\documentstyle[emulateapj]{article}
\newcommand\Ha{H$\alpha$}
\newcommand\ha{\rm H\alpha}
\newcommand\Lha{L_{\rm H\alpha}}
\newcommand\Lhacrit{L_{\rm H\alpha,crit}}
\newcommand\Sigha{\Sigma_{\rm H\alpha}}
\newcommand\mhi{$M_{\rm HI}$}
\newcommand\fwim{$f_{\rm WIM}$}

\def\Ha{H$\alpha$}
\newcommand\hi{\ion{H}{1}}
\newcommand\hii{\ion{H}{2}}
\newcommand\etal{et\thinspace al.~}

\newcommand\msol{\rm\,M_\odot}

\newcommand\ergs{{\rm\,erg\,s^{-1}}}
\newcommand\kms{{\rm\,km\,s^{-1}}}
\newcommand\hipass{{\sc HiPASS}}
\newcommand\hiiphot{HIIphot}

\def\spose#1{\hbox to 0pt{#1\hss}}
\def\simpropto{\mathrel{\spose{\raise 3pt\hbox{$\propto$}}
     \lower 3.0pt\hbox{$\sim$}}}
\def\lta{\mathrel{\spose{\lower 3pt\hbox{$\mathchar"218$}}
     \raise 2.0pt\hbox{$\mathchar"13C$}}}
\long\def\symbolfootnote[#1]#2{\begingroup%
\def\thefootnote{\fnsymbol{footnote}}\footnote[#1]{#2}\endgroup}

\received{24 July 2006}
\accepted{}
\slugcomment{Accepted 28-February-2007 to the Astrophysical Journal}

\lefthead{M. S. Oey et al.}
\righthead{Warm Ionized Medium and Escaping Radiation}

\begin{document}

\title{
% {\rm\small *****
% DRAFT version \today\ --- please do not circulate
%   Resubmitted version
% ***** } \\ \bigskip
% Diffuse, Warm Ionized Medium in the SINGG Survey of \hi-Selected Galaxes
The Survey for Ionization in Neutral-Gas Galaxies:  \break III.  Diffuse,
Warm Ionized Medium and Escape of Ionizing Radiation
}
\author{M. S. Oey\altaffilmark{1}, 
G. R. Meurer\altaffilmark{2},
S. Yelda\altaffilmark{1,3},
E. J. Furst\altaffilmark{4},
S. M. Caballero-Nieves\altaffilmark{1,5},
D. J. Hanish\altaffilmark{2},
E. M. Levesque\altaffilmark{6},
D. A. Thilker\altaffilmark{2},
G. L. Walth\altaffilmark{7},
J. Bland-Hawthorn\altaffilmark{8}
M. A. Dopita\altaffilmark{9},
H. C. Ferguson\altaffilmark{11},
T. M. Heckman\altaffilmark{2},
M. T. Doyle\altaffilmark{10},
M. J. Drinkwater\altaffilmark{10},
K. C. Freeman\altaffilmark{9},
R. C. Kennicutt, Jr.\altaffilmark{12},
V. A. Kilborn\altaffilmark{13,15},
P. M. Knezek\altaffilmark{14},
B. Koribalski\altaffilmark{15},
M. Meyer\altaffilmark{11},
M. E. Putman\altaffilmark{1},
E. V. Ryan-Weber\altaffilmark{12},
R. C. Smith\altaffilmark{17},
L. Staveley-Smith\altaffilmark{15},
R. L. Webster\altaffilmark{18},
J. Werk\altaffilmark{1},
M. A. Zwaan\altaffilmark{16},
}

\altaffiltext{1}{University of Michigan, Department of Astronomy, 830 Dennison
	Building, Ann Arbor, MI\ \ \ 48109-1042}
\altaffiltext{2}{Johns Hopkins University, Department of Physics and Astronomy, 
        3400 N. Charles St., Baltimore, MD\ \ \  21218-2686}
\altaffiltext{3}{Present address:  University of California,
	Department of Physics and Astronomy, P. O. Box 951547, Los
	Angeles, CA\ \ \ 90095-1547}
\altaffiltext{4}{Participant in the Research Experience for Undergraduates program,
        Northern Arizona University; Bucknell University, Department
	of Physics, Lewisburg, PA\ \ \ 17837; Present address:  344
	Greenlow Rd., Catonsville, MD\ \ \ 21228}
\altaffiltext{5}{Present address:  Georgia State University,
	Department of Physics and Astronomy, P. O. Box 4106, Atlanta,
	GA\ \ \ 30303}
\altaffiltext{6}{Massachusetts Institute of Technology, Department of
	Physics, 77 Massachusetts Ave., Cambridge, MA\ \ \ 02139;
        Present address:  Institute for Astronomy, 2680 Woodlawn Dr.,
        Honolulu, HI\ \ \ 96822}
\altaffiltext{7}{Observatories of the Carnegie Institution of Washington, 
        813 Santa Barbara St., Pasadena, CA\ \ \ 91101}
\altaffiltext{8}{Anglo-Australian Observatory, P. O. Box 296, Epping,
	NSW 2121, Australia}
\altaffiltext{9}{Australian National University, Research School of
	Astronomy and Astrophysics, Cotter Road, Weston Creek, ACT
	2611, Australia}
\altaffiltext{10}{University of Queensland, Department of Physics,
	St. Lucia, QLD 4072, Australia}
\altaffiltext{11}{Space Telescope Science Institute, 3700 San Martin
	Dr., Baltimore, MD\ \ \ 21218}
\altaffiltext{12}{Institute of Astronomy, University of Cambridge,
	Madingley Road, Cambridge, CB3 0HA, United Kingdom}
\altaffiltext{13}{Swinburne University of Technology, Centre for
	Astrophysics and Supercomputing, Mail 39, P. O. Box 218,
	Hawthorn, VIC 3122, Australia}
\altaffiltext{14}{WIYN, Inc., 950 N. Cherry Ave., Tucson, AZ\ \ \ 85721}
\altaffiltext{15}{Australia Telescope National Facility, CSIRO,
	P. O. Box 76, Epping, NSW 1710, Australia}
\altaffiltext{16}{European Southern Observatory,
	Karl-Schwarzschild-Str. 2, D-85748 Garching b. M\"unchen, Germany}
\altaffiltext{17}{Cerro Tololo Inter-American Observatory, Casilla
	603, La Serena, Chile}
\altaffiltext{18}{University of Melbourne, School of Physics,
	Parkville, VIC 3010, Australia}

\begin{abstract}
We use the first data release from the SINGG \Ha\ survey of
\hi-selected galaxies to study the quantitative behavior of the diffuse,
warm ionized medium (WIM) across the range of properties represented
by these 109 galaxies.  The mean fraction \fwim\ of diffuse ionized gas in
this sample is $0.59\pm 0.19$, slightly higher than found in previous
samples.  Since lower surface-brightness galaxies tend to have higher
\fwim, we believe that most of this difference is due to selection
effects favoring large, optically-bright, nearby galaxies with high
star-formation rates.  As found in previous studies, there 
is no appreciable correlation with Hubble type or total star-formation
rate.  However, we find that starburst galaxies, defined here
by an \Ha\ surface brightness $> 2.5\times 10^{39}\ \rm erg\ s^{-1}\
kpc^{-2}$ within the \Ha\ half-light radius, do show much lower
fractions of diffuse \Ha\ emission.  The cause apparently is not dominated by
a lower fraction of field OB stars.  However, it is qualitatively
consistent with an expected escape of ionizing radiation above a threshold
star-formation rate, predicted from our model
in which the ISM is shredded by pressure-driven supernova feedback.
The \hi\ gas fractions in the starburst galaxies are also lower,
suggesting that the starbursts are consuming and ionizing all the gas,
and thus promoting regions of density-bounded ionization.  If true,
these effects imply that some amount of Lyman continuum radiation is
escaping from most starburst galaxies, and that WIM properties and
outflows from mechanical feedback are likely to be pressure-driven.
However, in view of previous studies showing that the escape
fraction of ionizing radiation is generally low, it is likely that
other factors also drive the low fractions of diffuse ionized gas in
starbursts. 
\end{abstract}


\section{Introduction}

Ionizing radiation is one of the three major feedback processes
from massive stars, alongside mechanical and nucleosynthetic effects,
that are propagated by the most massive stellar
populations.  The hot temperatures and powerful luminosities of these
stars yield prodigious emission rates of H-ionizing photons ($48
\lta \log Q_0/{\rm s^{-1}} \lta 50$), thereby producing
luminous regions of ionized gas that are visible at great distances.

Because of the high ionizing luminosities, ranging up to $\sim 3\times
10^{38}\ \ergs$ for individual stars, radiative feedback is
energetically important and has fundamental consequences for the
evolution of the interstellar medium (ISM) of host galaxies and the
surrounding intergalactic medium (IGM).  The warm ($10^4$ K) component
of the multiphase ISM, which is the most massive component of ionized
gas in galaxies (Walterbos 1998), is thought to 
result primarily from radiative feedback, and the resulting ISM phase
balance strongly affects evolutionary processes like star formation
and ISM gas dynamics.  Furthermore, of intense current interest,
the escape fraction and energies of ionizing photons from galaxies 
are crucial to the ionization state of the
IGM and reionization of the early universe.  With current rapid
advances in absorption-line and emission-line probes of
the high-redshift Universe, we are urgently in need of a quantitative
understanding of radiative feedback processes to interpret cosmic history
and galaxy formation.

The total \Ha\ emission from star-forming galaxies is divided roughly
in half, between classical, discrete \hii\ regions, and the diffuse, warm
ionized medium (WIM), often also referred to as diffuse ionized gas
(DIG).  The WIM is generally thought to be ionized by
massive stars, too, since they are the only candidate capable of 
providing the necessary power (Reynolds 1984).  However, the
ionization and energetics of the WIM are poorly understood.  For
example, does the WIM always comprise around 40\% of the total \Ha\
emission in star-forming galaxies, as suggested by the extant
observations (e.g., Wang et al. 1999; Hoopes {\etal}1996; Ferguson
{\etal}1996)?  If so, why would the WIM fraction be independent of 
galaxy parameters like star formation rate and Hubble type?  It is
thought that roughly half of the diffuse ionization originates from
Lyman continuum radiation escaping from ordinary \hii\ regions (e.g., Oey \&
Kennicutt 1997), and half from massive stars in the field (e.g., Oey et
al. 2004; Hoopes \& Walterbos 2000).  While there is a spatial
correlation between \hii\ regions and diffuse emission (e.g., Zurita
et al. 2002; Ferguson et al. 1996; Hoopes et al. 1996),
Dopita et al. (2006) also
suggest that about half of the WIM simply may consist of extremely
evolved, filamentary \hii\ regions that may be difficult to detect as
such, and therefore are assigned to the WIM.
It also appears that photoionization by OB stars cannot
exclusively explain the ionization state of the WIM (e.g., Collins \&
Rand 2001; Reynolds {\etal}1999).  Thus, a clearer understanding of
the ionization processes and radiative transfer is needed to
understand the true role of massive stars.
And, if ionizing radiation escapes from \hii\ regions to ionize the
WIM, then does it also escape from galaxies altogether, thereby
affecting the ionization state of the IGM, as
predicted by, e.g., Clarke \& Oey (2002)?  If this occurs, then how
commonplace is it, and under what conditions does it happen? 

The first, necessary step is to empirically quantify the WIM
properties across all classes of star-forming galaxies, which is
now possible with a new, definitive dataset:  the
Survey for Ionization in Neutral-Gas Galaxies (SINGG; Meurer
{\etal}2006; hereafter Paper~I).  SINGG is an \Ha\ and $R$-band  imaging survey of 468
galaxies selected only on the basis of their \hi\ emission from the
\hi\ Parkes All-Sky Survey (\hipass; Barnes 2001).  Since \hipass\ is
a blind \hi\ survey of the entire southern sky within radial velocity
$12,700\ \kms$, the detected galaxies span essentially the entire range of 
star formation properties that occur, given an adequate gas supply.
SINGG consists of a subsample of the \hipass\
galaxies, up to radial velocities of $10,000\ \kms$, and uniformly
sampling the \hi\ mass range $7.0 < \log (M_{\rm HI}/\msol) < 11.0$.
It is therefore possible to investigate the global radiative feedback
effects essentially {\it across a complete parameter space of gas-rich
galaxy properties.} 

With the SINGG dataset, we can quantify parameters like the fraction of
total \Ha\ luminosity occupied by the WIM, and
its relation to galaxy properties like star-formation rate, \hi\ mass,
and stellar luminosity.  The patterns that emerge
from the data will clarify the physical processes associated
with this major component of the ISM.

\section{\Ha\ Data Analysis}

We adopt the common definition of the WIM, namely, that it is the diffuse \Ha\
emission in excess, and outside of, the classical \hii\ regions (e.g.,
Hoopes et al. 1996).  Since there are many ways of defining \hii\
region boundaries, any quantitative results on the WIM parameters
will necessarily depend on the methods used, and so quantitative
results from any study of the WIM should be treated with appropriate
caution.  Nevertheless, as we demonstrate below, our general
qualitative results are robust.  Future studies incorporating an
additional parameter to define the WIM, such as kinematic line widths,
may yield more robust and useful definitions of the WIM (Shopbell \&
Bland-Hawthorn 1998). 

We report here on results from 109 galaxies in the SINGG Release 1
(SR1) subset (Paper~I).  These generally
correspond to a range in \hi\ mass of $7.5 < \log (M_{\rm HI}/\msol) <
10.6$, although some targets turned out to have multiple galaxies in
the field of a single \hi\ detection, and so for these
siblings we have only an upper limit on \mhi.  The galaxies have
distances of 4 -- 73 Mpc, with most in the range 10 -- 20 Mpc
(see Paper~I; Hanish et al. 2006).  Preliminary data reduction and flux
calibration for the sample were carried out by the SINGG pipeline (see Paper~I),
including Galactic and internal extinction corrections, and corrections
for [\ion{N}{2}] inclusion in the \Ha\ filter bandpasses.

An important parameter for this sample of \Ha\ observations is the
star formation per unit area, or star-formation intensity
(SFI\symbolfootnote[1]{Paper~I abbreviated the star-formation per unit
area as `SFA.'}).  For much of the work that follows, we adopt
three bins of SFI, corresponding to the effective \Ha\ surface
brightness $\Sigha$ within the star-forming disks, which is given by:
\begin{equation}\label{eq_sfi}
\Sigha = \frac{L_{\ha}}{2\pi R_{e,\ha}^2} \quad ,
\end{equation}
where $R_{e,\ha}$ is the effective (half-light) radius in \Ha\ from
Paper~I.  We refer to galaxies having $38.4 <
\log\Sigha \leq 39.4$ as ``normal,'' those having $\log\Sigha >
39.4$ as ``starburst,'' and those having $\log\Sigha \leq 38.4$ as
``sparse.''  This binning, in units of $\rm \log erg\ s^{-1}\
kpc^{-2}$, was chosen to assign galaxies
whose star-forming disks are mostly packed with merging \hii\
regions as ``starbursts.''  This is much more generous
than Heckman's (2005) definition in terms of SFI;
using the conversion from Kennicutt et al. (1994) to relate SFR and
\Ha\ luminosity: 
\begin{equation}\label{eq_sfrha}
\rm SFR\ (\msol\ yr^{-1}) = {\it L}_{H\alpha}\ (erg\ s^{-1})\ /\
    1.26\times10^{41} \quad ,
\end{equation}
our criterion of $\Sigha > 2.5\times 10^{39}\ \rm erg\ s^{-1}\
kpc^{-2}$ corresponds to 0.02 $\msol\ 
yr^{-1}\ kpc^{-2}$, whereas Heckman gives a range of 1 -- 100 $\msol\
yr^{-1}\ kpc^{-2}$ for typical starbursts.  However, that definition
is largely based on small portions of galaxies containg a starburst
(Meurer et al. 1997).  Other common starburst definitions include
EW$_{50}$(\Ha) $>$ 50 \AA, as in Paper~I, which refers to the
equivalent width within the \Ha\ half-light radius.  Our adopted
starburst definition based on $\Sigha$ is more appropriate to the global
properties used here:  Meurer (2007, in preparation) shows that a
single threshold of $\Sigha \gtrsim 39.4$ is sufficient to
isolate the well-known starbursts in the SR1 sample, and is
``cleaner'' than an equivalent EW$_{50}$(\Ha) cut.
% There are other
% definitions of starbursts, for example, the one adopted in Paper~I:
% $\rm EW^\prime_{50}(\ha)\geq50$ \AA,  
% Figure~\ref{f_ewsfi} shows the relation between $\Sigha$ and
% EW$^\prime_{50}(\ha)$.  There is a large scatter, and these
%definitions do not agree cleanly, although they do cover a similar
% parameter space.  
We therefore feel that an \Ha\ surface
brightness, or intensity criterion suggested by Heckman (2005) is
intuitively the most appropriate, but readers should consider
differences in definitions when considering the literature on starbursts.
We also define a category of  ``nuclear starbursts,''
which are galaxies whose star formation is strongly dominated by the
nuclear region; these are visually assigned.
Figure~\ref{f_categories} shows example \Ha\ images of each star
formation category.

Similar to the analysis by Wang et al. (1999),
Figure~\ref{f_sbdist}$a$ shows the \Ha\ surface brightness 
distributions in \Ha\ flux per pixel $S_{\ha}$, after smoothing the
\Ha\ images with a $9\times9$ pixel (3.87\arcsec\ diameter) box car.
% for eight representative galaxies.  As done by Wang et al. (1999),
We include only pixels showing emission above a detection limit of
$2\sigma$ above the background level in each image, which was derived
as discussed in Paper~I.   
% with the background level taken to be the statistical mode of the image.
The galaxies are
color-coded according to $\Sigha$, with starbursts at the red extreme,
and the lowest $\Sigha$ at the violet and black extreme.
% For the units of
% \Ha\ surface brightness, we use the median $S_{\ha}$, which is
% somewhat less sensitive to the total SFR, instead of the mean value,
% which was used by Wang et al.  
% The galaxies shown in Figure~\ref{f_sbdist}$a$ are the four shown in
% Figure~\ref{f_categories}, plus four additional galaxies:
% J0216--11, J0351--38, J0459--26 and J1339--31.
% The ``normal'' and ``sparse'' galaxies are plotted with solid
% lines and the ``starbursts'' with dashed lines.  Figure~\ref{f_sbdist}$b$ 
% shows the average co-added \Ha\ surface brightness distributions for
% ``sparse,'' ``normal,'' and  ``starburst'' galaxies with the dotted,
% solid, and dashed lines, respectively.  
It is apparent that the starbursts show significantly flatter slopes
at lower $S_{\ha}$, indicating that they have smaller relative
contributions of diffuse \Ha\ emission compared to the \hii\ regions.
We determined the slopes $\alpha$ using a
least-squares fit to the logarithmic quantities plotted in
Figure~\ref{f_sbdist}$b$, within the range $-1.5 < \log(S_{\rm
H\alpha}/S_{\rm H\alpha,eff}) < 0.0$.  The normalizing
surface brightness $S_{\rm H\alpha,eff}$ corresponds to the isophote
that includes half the total \Ha\ flux of each object.
% We fitted the slopes $\alpha$ to the logarithmic quantities 
% % \footnote{Note that the corresponding exponent on the linear
% % quantities is $\alpha + 1$.}  (THIS ISN'T LOG A (LOG B) VS LOG B)
% plotted in Figure~\ref{f_sbdist}$b$, within the range 
% $0.1<\log(S_{\ha}/S_{\ha,eff}) < 1.0$, showing
% for most galaxies; 13 galaxies
% were fitted within the range $1.25<\Sigha/\Sigma_{\rm H\alpha,detect}< 3.81$
% because of the limited range in their surface brightness
% distributions.  J1131--02:S3 has almost no star formation and its
% distribution could not be fitted.
% the fitted slopes for the average
% logarithmic $S_{\ha}$ distribution for the ``sparse,'' ``normal,'' and
% ``starburst'' galaxies are, respectively, 
% % $-2.0\pm 0.9,\ -1.4\pm 0.4,$ and $-0.8\pm 0.4$.  The fitted slopes
% of the co-added distributions for these respective categories are
% $-2.14\pm 0.07,\ -1.33\pm 0.03$, and $-0.79\pm 0.02$. 
% MSO FINAL:  $-1.29\pm 0.06,\ -1.00\pm 0.05$, and $-0.36\pm 0.05$.
% Thus, we see that the surface brightness distributions confirm the lower
% \Ha\ diffuse fractions for our starburst galaxies.
%%  Figure~\ref{f_sbslopesfi} shows
Figure~\ref{f_sbdist}$b$ shows
these fitted slopes as a function of $S_{\ha}$, color-coded as before,
again demonstrating the trend of flattening slope with increasing SFI
for the whole sample.

\begin{figure*}
\epsscale{1.0}
% \epsscale{1.3}
%\vspace*{2.0in}
%\hspace*{-3.5in}
\plotone{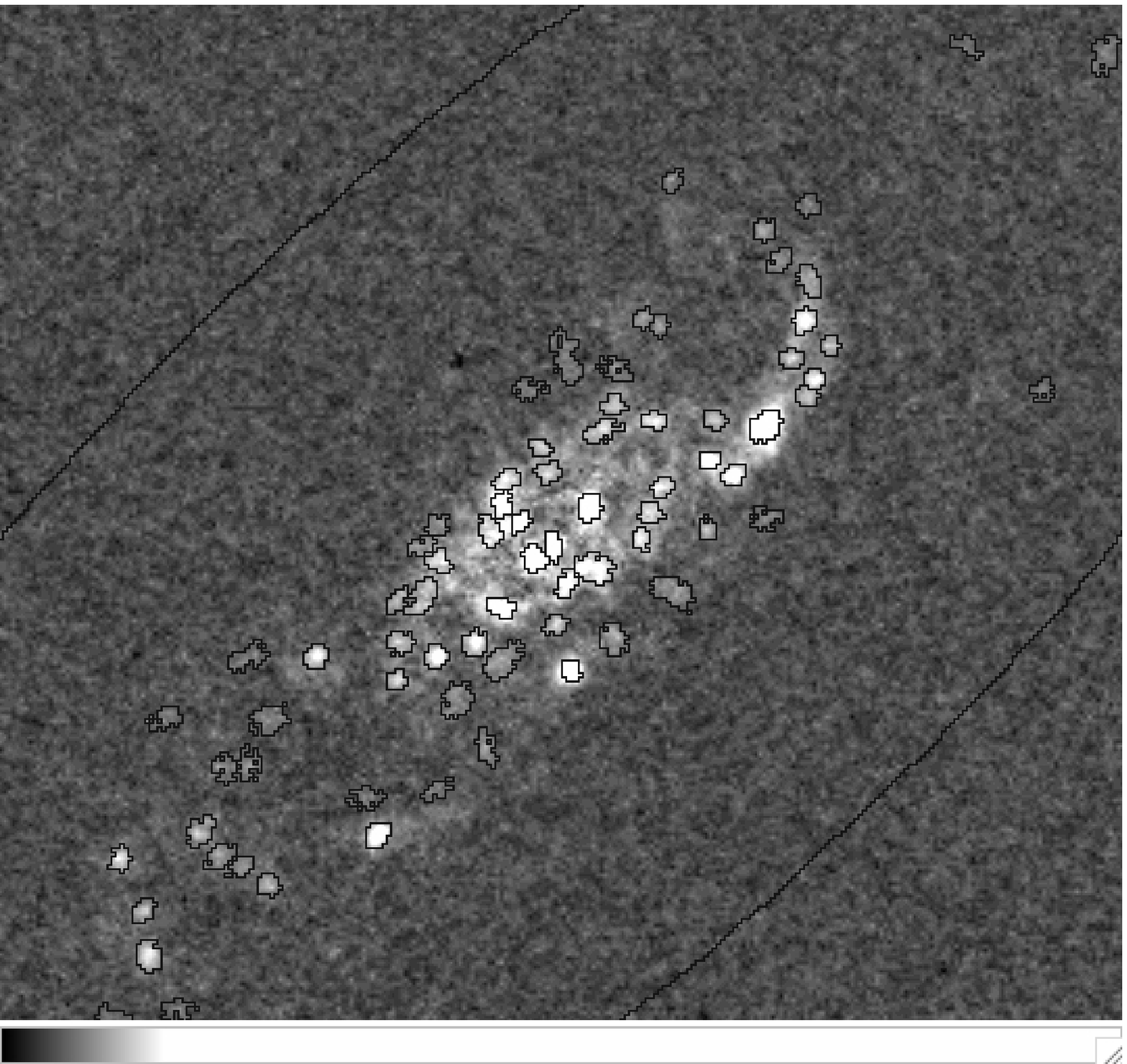}
% \plotone{J0412.eps}
 \plotone{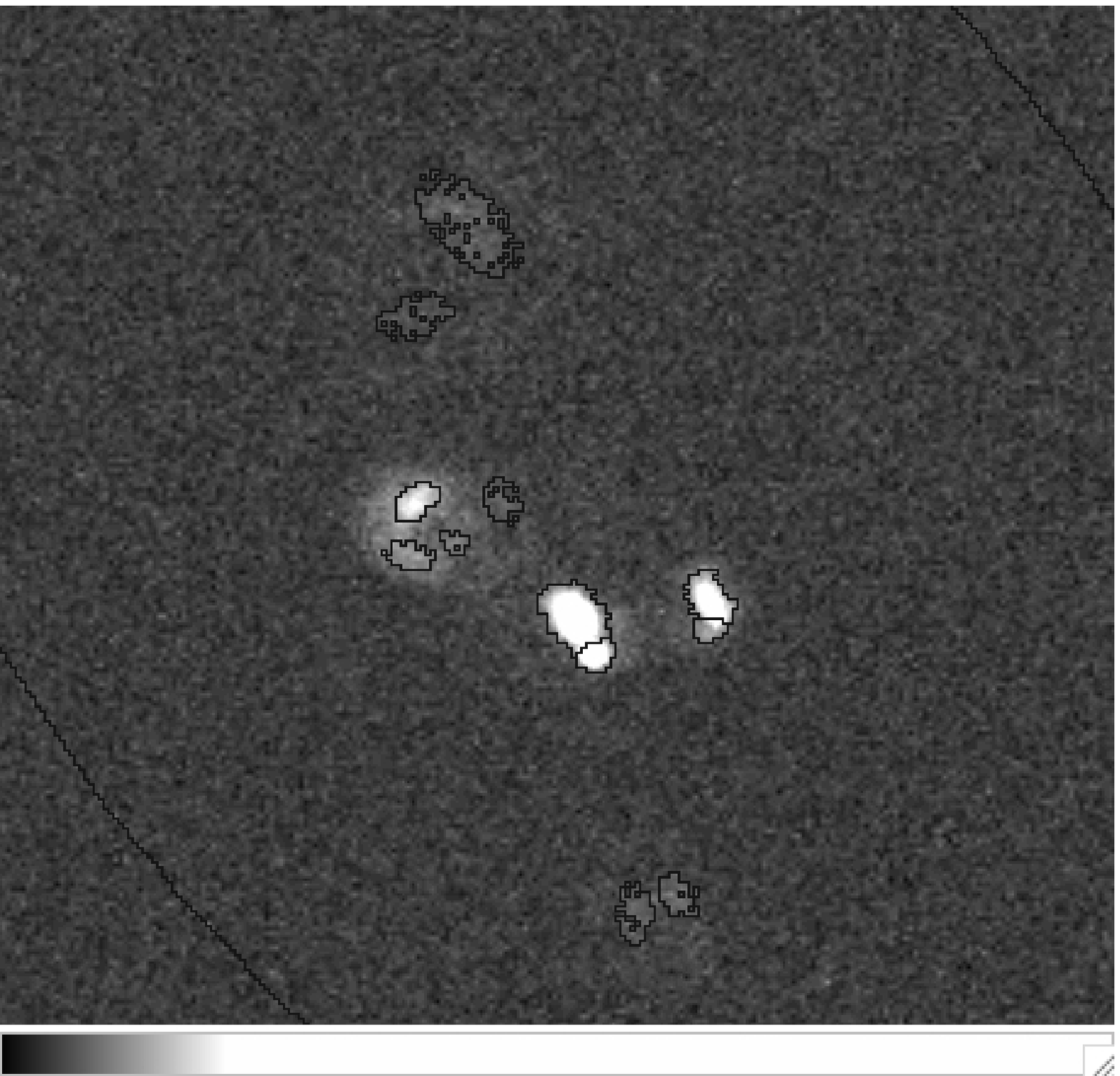}
%\plotone{J0031.eps}
%\vspace*{-4.5in}
\caption{Representative examples of the galaxy categories.  ($a$)
  ``Normal:'' J0412+02, $\log\Sigha =38.92$; ($b$) ``Sparse:''
  J0031--22, $\log\Sigha =38.26$;  ($c$) ``Starburst:'' J0355--42,
  $\log\Sigha=39.40$; ($d$) ``Nuclear Starburst:'' J0209--10:S2.
  $\Sigha$ is quoted in units of $\rm erg\ s^{-1}\ kpc^{-2}$.
  {\sc Hii} region boundaries defined by \hiiphot\ are outlined in black.
  The large elliptical apertures indicated by the black lines around
  the galaxies are those defined from $R$-band images by Paper~I (their
  $r_{\rm max}$) for the total galaxy flux measurements.  The images
  are roughly 1.8$^\prime$ square, with north up and east to the left;
  all are displayed with the same gray scale.
\label{f_categories} }
\end{figure*}

\setcounter{figure}{0}

\begin{figure*}
\epsscale{1.0}
%\epsscale{0.5}
%\vspace*{2.0in}
%\hspace*{-3.5in}
\plotone{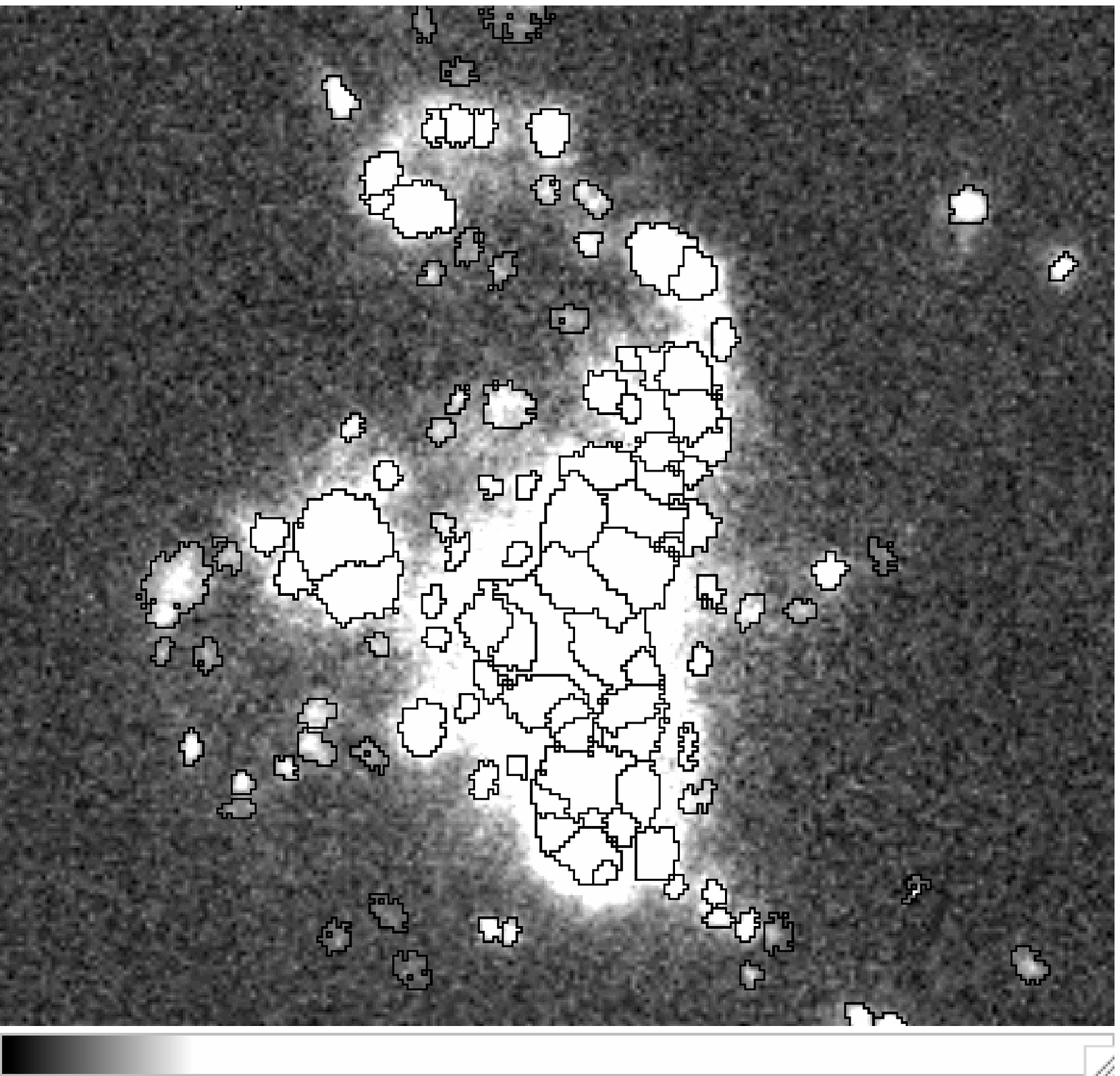}
% \plotone{J0355.eps}
\plotone{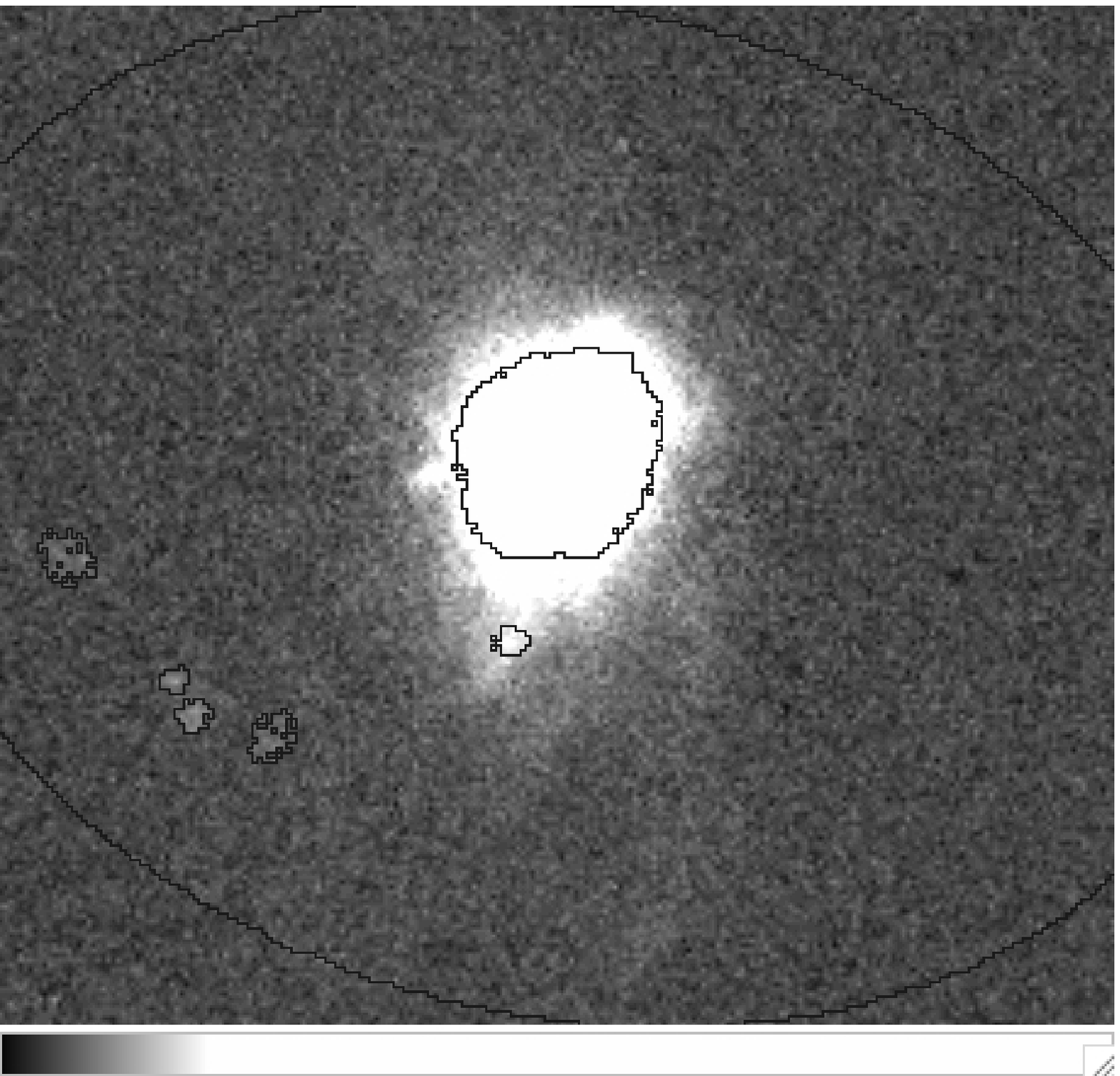}
% \plotone{J0209.eps}
%\vspace*{-4.5in}
\caption{continued}
\end{figure*}

\begin{figure*}
%\epsscale{1.0}
\epsscale{2.0}
% \vspace*{2.5in}
%\hspace*{-1.0in}
\plotone{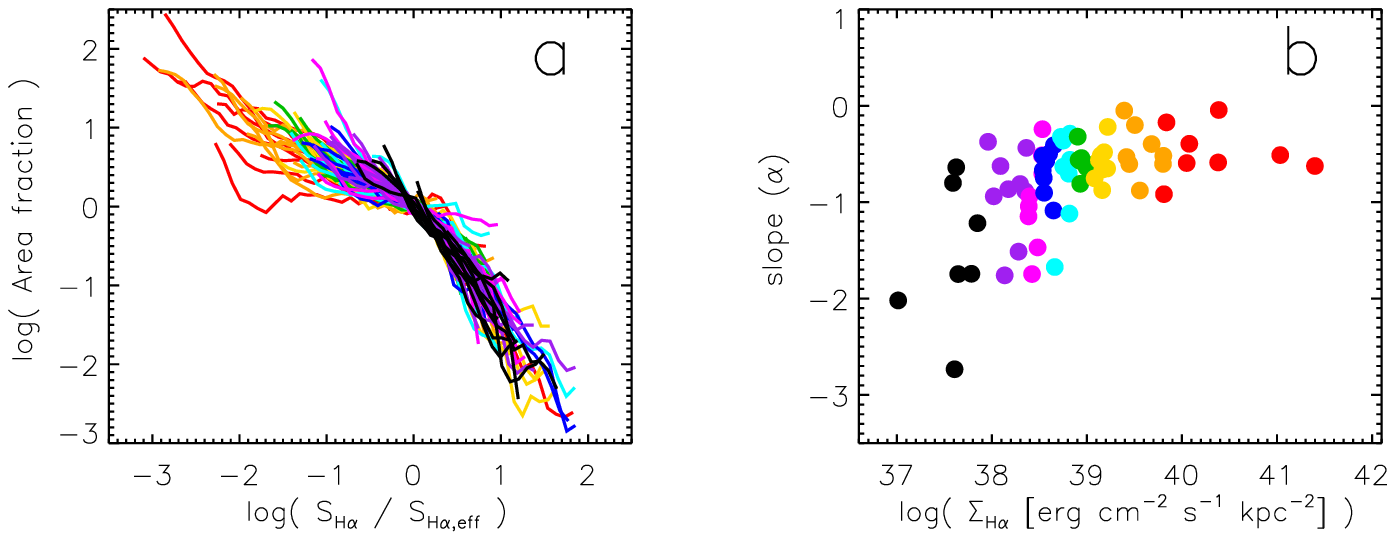}
% \plotone{replfig2_rev_fixed.eps}
%\hspace*{0.5in}
% \plotone{f2b.eps}
% \plotone{sbdistb.ps}
% \vspace*{-2.0in}
\caption{Differential \Ha\ surface brightness distributions for the
  individual galaxies (Figure~\ref{f_sbdist}$a$),
  color-coded according to $\Sigha$;
  starburst galaxies are plotted in red, and galaxies with the lowest
  $S_{\ha}$ in violet and black.  The distributions are shown
  as fractional area of all pixels above the detection threshold, and
  $S_{\ha}$ is normalized to $S_{\ha,eff}$, the value above which
  the distribution accounts for half the \Ha\ flux.  The surface
  brightness distributions are also normalized at this value.
  Figure~\ref{f_sbdist}$b$ shows the slopes of the low-luminosity end
  of the \Ha\ surface brightness distributions as a function of mean
  total $\Sigha$. 
\label{f_sbdist} }
\end{figure*}

% \begin{figure*}
% \epsscale{1.0}
% % \vspace*{2.5in}
% %\hspace*{-1.0in}
% % \plotone{f2a.eps}
% \plotone{sbdista.ps}
% %\hspace*{0.5in}
% % \plotone{f2b.eps}
% \plotone{sbdistb.ps}
% \vspace*{-2.0in}
% \caption{Differential \Ha\ surface brightness distributions for eight
%   individual galaxies (Figure~\ref{f_sbdist}$a$, see text), with
%   starburst galaxies shown by dashed lines and other galaxies by solid lines.
%   Figure~\ref{f_sbdist}$b$ shows the co-added average distributions for
%   galaxies categorized as ``sparse'' (dotted line), ``normal'' (solid
%   line), and ``starburst'' (dashed line).  
% %  There is more noise for
% %  points where fewer galaxies contribute to the average value.
%   The distributions are shown
%   as fractional area of all pixels above the detection threshold, and
%   $S_{\ha}$ is given in units of the median value (see text).  
% \label{f_sbdist} }
% \end{figure*}
% 
% \begin{figure*}
% \epsscale{1.5}
% %\vspace*{2.0in}
% %\hspace*{-3.5in}
% % \plotone{f?.eps}
% \plotone{sbslope_sfi.ps}
% %\vspace*{-4.5in}
% \caption{Slopes of the low-luminosity end of the \Ha\ surface
%   brightness distributions as a function of mean total $\Sigha$.  The
%   vertical dashed lines show our boundaries between ``sparse,''
%   ``normal,'' and ``starburst'' galaxies.  Symbols with black dots
%   denote nuclear starbursts, as before.
% \label{f_sbslopesfi} }
% \end{figure*}
 
We can examine the conventional fraction \fwim\ of diffuse \Ha\
emission by defining the spatial division between the WIM and \hii\
regions, as is commonly done.
Because the diffuse \Ha\ emission of the WIM is the emission in excess,
and outside of, the classical \hii\ regions, defining the
WIM therefore requires defining the \hii\ regions as
well.  Since photometry of irregular, extended \hii\ regions is tricky,
we compared some of the different algorithms available for
automated \hii\ region identification and photometry (e.g., McCall
{\etal}1996; Knapen 1998; Thilker {\etal}2000).  The \break \hiiphot\
software by Thilker {\etal}(2000) yields reliable nebular catalogs, and
we opted to use this code, based on the \hii\ region definition
criteria and user control over vital parameters.  \hiiphot\ then
identifies objects above a threshold signal-to-noise, and determines the
boundaries according to a user-defined, limiting radial gradient in
\Ha\ surface brightness, known as the
``terminal gradient''.  \hii\ region luminosities are computed 
within the boundaries, subtracting the local background, and
the remaining diffuse \Ha\ background and exterior emission is defined
to be the WIM.  The background emission for the nebular regions is
computed from a two-dimensional interpolation scheme described by
Thilker et al. (2000).  We modified the original version of the code to
calculate the total WIM luminosity and fraction of the \Ha\ luminosity.
The \hiiphot\ algorithms are more fully described by Thilker et al. (2000),
and we also discuss them further below.

We first created blanking masks to exclude foreground stars and other
spurious features from the analysis.  The total galaxy apertures
were the same elliptical apertures defined by Paper~I, based
on the galaxy $R$-band isophotes.  In a few cases where the
galaxy's angular size exceeded that required by the standard
aperture-defining algorithm, we redefined the apertures by hand, using
the \hiiphot\ interface for that purpose.
Our independent measurements agree
well; the mean ratio of galaxy \Ha\ luminosities measured
by Paper~I, to those measured in this work, is
$\Lha(\rm Paper~I)/\Lha(\hiiphot)=1.07\pm0.29$.

\begin{figure*}
\epsscale{1.3}
%\vspace*{2.0in}
%\hspace*{-3.5in}
\plotone{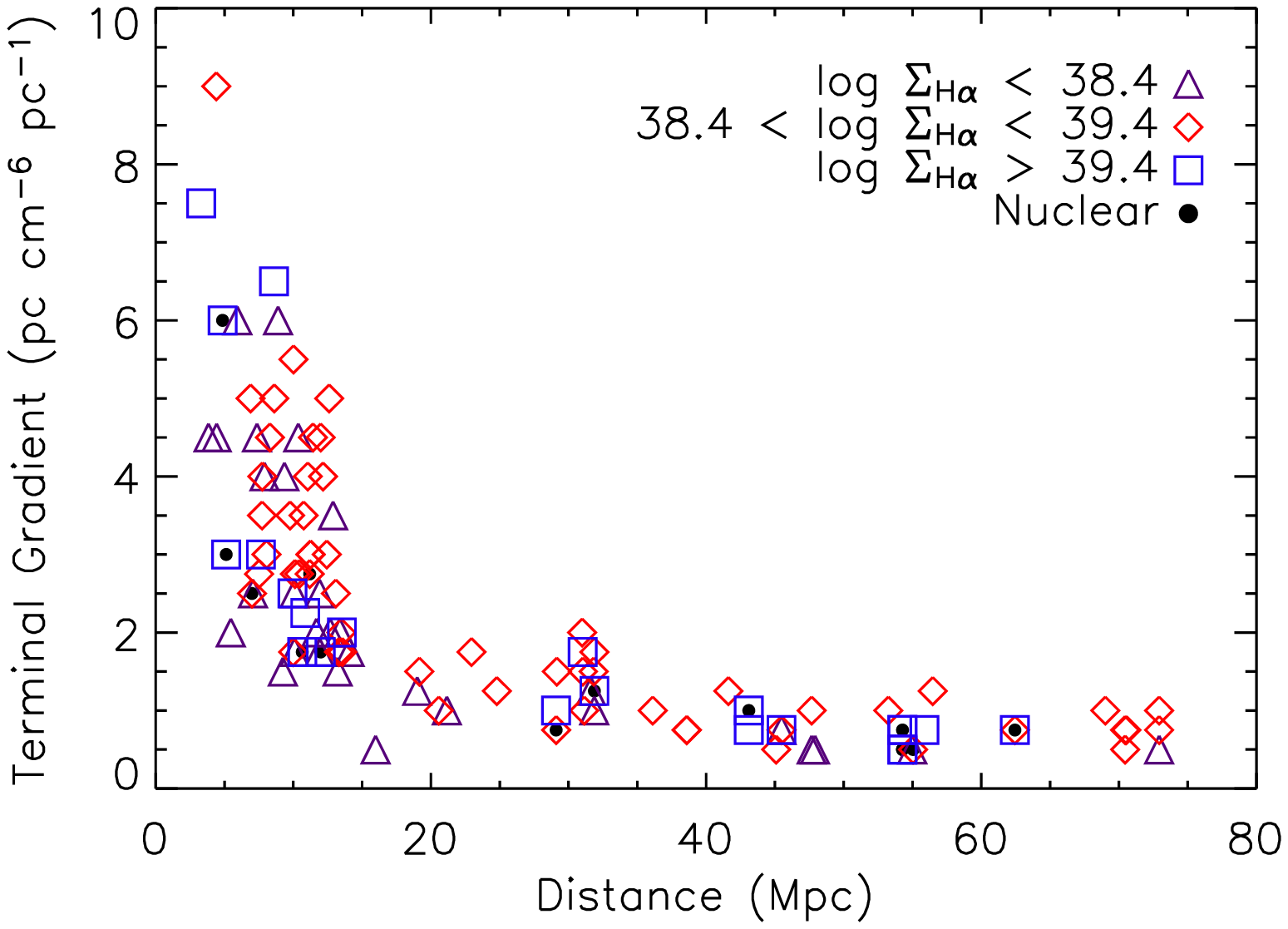}
% \plotone{termgrad_dist.ps}
%\vspace*{-4.5in}
\caption{Adopted \hiiphot\ terminal gradients vs distance.  The symbols show
  star-formation intensity, as measured by the \Ha\ surface brightness
  $\Sigha$, as shown.  Galaxies dominated by nuclear star formation
  are indicated by the central black dots.
\label{f_termgrads} }
\end{figure*}

\begin{figure*}
%\epsscale{1.0}
%\vspace*{2.0in}
% \hspace*{-3.5in}
\plotone{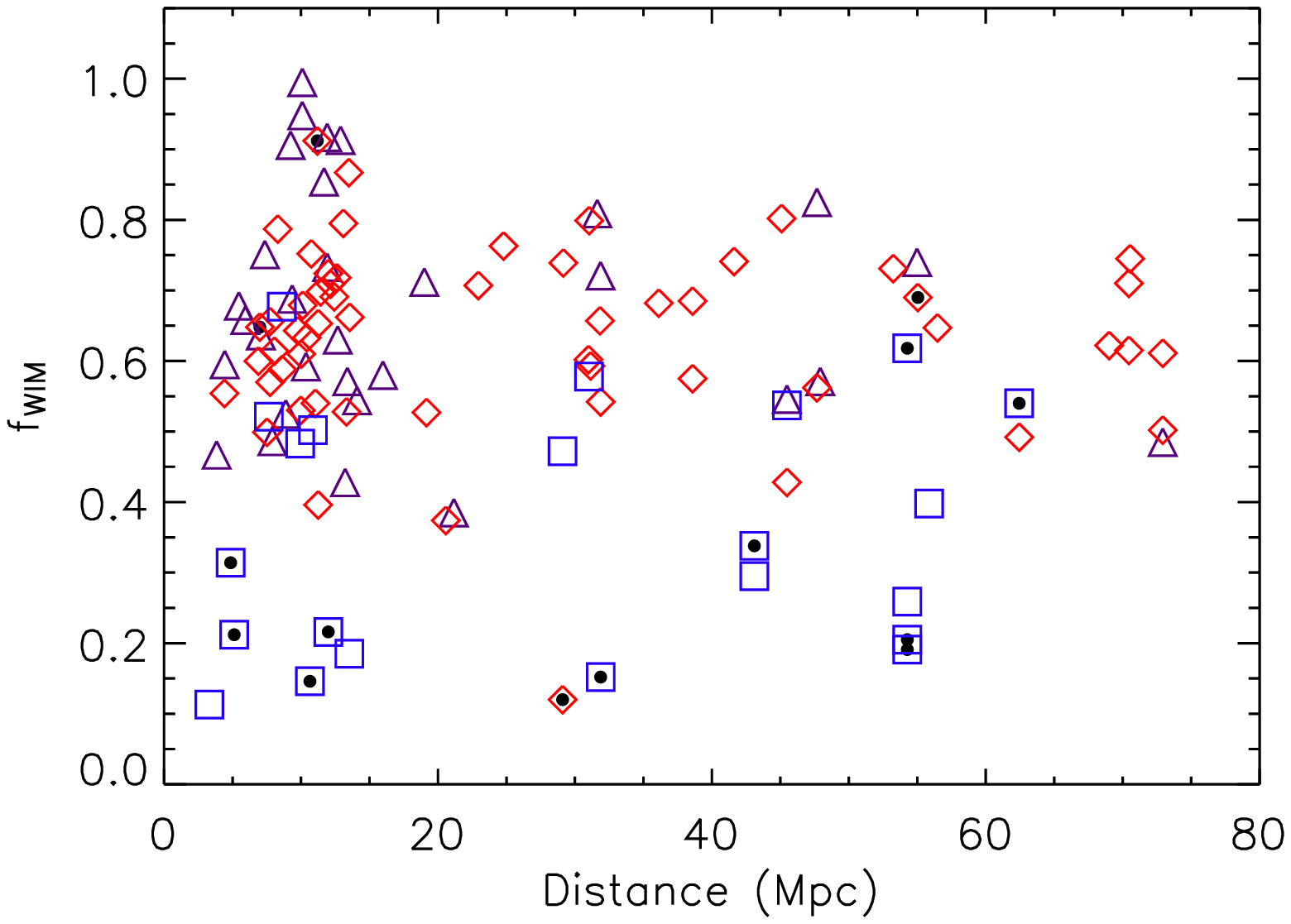}
% \plotone{diff_dist.ps}
% \plotone{diff_distppp.ps}
%\vspace*{-4.5in}
\caption{\Ha\ fraction of diffuse, warm ionized medium vs
  galaxy distance.  
%  Figure~\ref{f_diffdist}$a$ shows \fwim\ derived by the \hiiphot\
%  algorithm, and Figure~\ref{f_diffdist}$b$ shows \fwim\ derived using
%  the PPP method.
  Symbols are as in Figure~\ref{f_termgrads}.
\label{f_diffdist} }
\end{figure*}

The \hiiphot\ terminal gradient criterion for defining the \hii\
region boundaries depends on seeing conditions and distance, because
classical \hii\ regions have a strong drop in surface brightness at
the Str\"omgren edge.  Thus, for high spatial resolution in nearby
galaxies and/or good seeing, the terminal 
gradient that defines the object edge has a higher value than
for low resolution.  We therefore allocated the terminal gradients for
each galaxy based on visual inspection, with values ranging over
0.5 -- 9.0 pc $\rm cm^{-6}\ pc^{-1}$.  We find that the resulting
diffuse \Ha\ fractions \fwim\ are not highly sensitive to the
adopted terminal gradient; \fwim\ typically varies by $\lesssim 0.1$ for
termgrad variations of 25--50\%.  
Within a distance of $\sim$20 Mpc, we see the full range of terminal
gradients (Figure~\ref{f_termgrads}).  Beyond that distance, however, 
the terminal gradients converge to about 1 pc $\rm cm^{-6}\ pc^{-1}$,
again because the spatial resolution limits the \hii\ region boundary
definitions and resulting photometry.  As has been noted in the past
(e.g., Deharveng et al. 1988), the spatial resolution
affects \hii\ region luminosity functions for distant galaxies.
However, the total photometry of the \hii\ regions and WIM
apparently are not strongly affected, since \hiiphot\ computes a
diffuse background for the objects, and also because the WIM has a
larger scale height than the \hii\ regions, which lessens the
importance of line-of-sight confusion between \hii\ regions and WIM.
Figure~\ref{f_diffdist} shows the WIM fraction as a function of 
distance, and there is no apparent correlation, demonstrating that
effects due to spatial resolution are minimal.  We therefore include
the entire SR1 sample over the full distance range in our study of the
WIM behavior.  

% and indicated by the
% black dots in Figures~\ref{f_termgrads} and \ref{f_diffdist}.

Figure \ref{f_diffdist} shows that the starburst
galaxies have lower WIM fractions than the rest of the sample.  
% We again
% caution that these quantitative results for the starburst galaxies
% depend on the  methods and definitions used to distinguish the diffuse
% \Ha\ emission from the \hii\ regions.
Since the WIM is usually
defined simply as the remainder of a galaxy's \Ha\ emission outside of
the discrete \hii\ regions, the best way to measure its luminosity is
by careful photometry and subtraction of the \hii\ region luminosities
(e.g., Hoopes et al. 1996).  We believe that \hiiphot\ is currently
the best automated routine for 
nebular photometry because it carries out a detailed definition of the
\hii\ region boundaries, and it also accounts for varying background
emission levels (see above).  The latter are a problem for some
methods, like those that isolate the WIM by assigning it a fixed
surface brightness level; Thilker et al. (2002) provide detailed
discussion of various methods that have been used for defining diffuse
\Ha\ emission.

\begin{figure*}
%\epsscale{1.0}
%\vspace*{2.0in}
%\hspace*{-3.5in}
\plotone{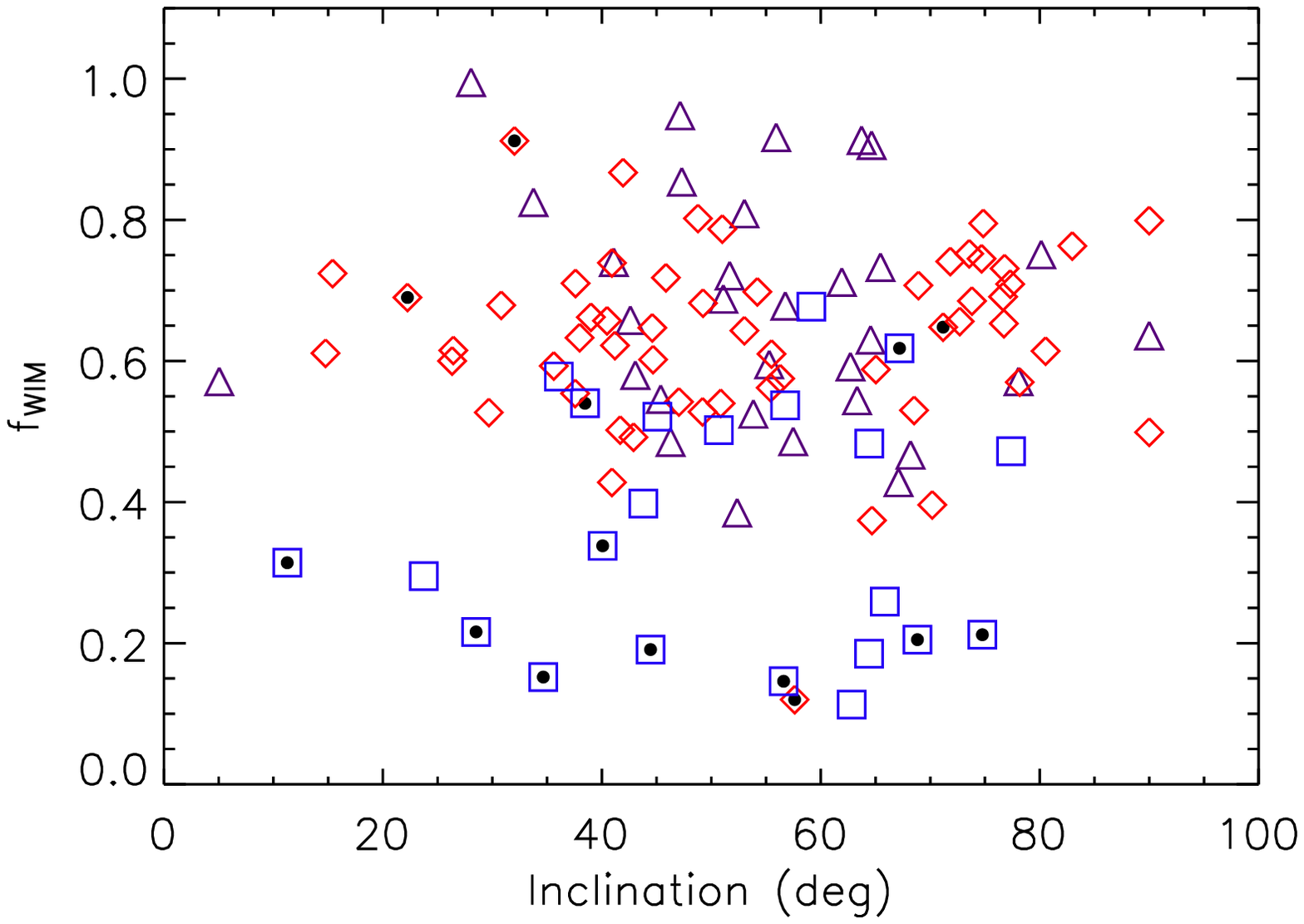}
% \plotone{diff_incl.ps}
%\vspace*{-4.5in}
\caption{WIM fraction vs galaxy angle of inclination.  Symbols for different
  star-formation categories are as in Figure~\ref{f_termgrads}.
\label{f_df_i} }
\end{figure*}

Figure~\ref{f_df_i} shows WIM fraction vs galaxy angle of inclination
$i$, with symbols as in Figure~\ref{f_termgrads}.  There are no
correlations, with all star-formation categories occurring at all
angles of inclination.  The measured \fwim\ also shows no
selection bias based on $i$.  Although photometry of the \hii\
regions, which are limited to the disk plane, becomes problematic at
large $i$ as the objects merge in the line of sight, photometry of the
WIM appears to be not strongly affected.  The \hiiphot\ algorithm, in
particular, minimizes inclination effects because it interpolates
local background values for \hii\ regions.  As described above, these
background values are included in the WIM photometry.  Our results are
consistent with those of Thilker et al. (2002), who also found no
apparent biases resulting from galaxy inclination.  Furthermore, the
WIM often has a large scale height relative to the classical \hii\
regions (e.g., Rossa \& Dettmar 2003; Collins \& Rand 2001; Veilleux
et al. 1995), which will tend to counteract inclination effects.  

\begin{figure*}
\epsscale{1.0}
% \epsscale{0.4}
%\vspace*{2.0in}
%\hspace*{-3.5in}
\plotone{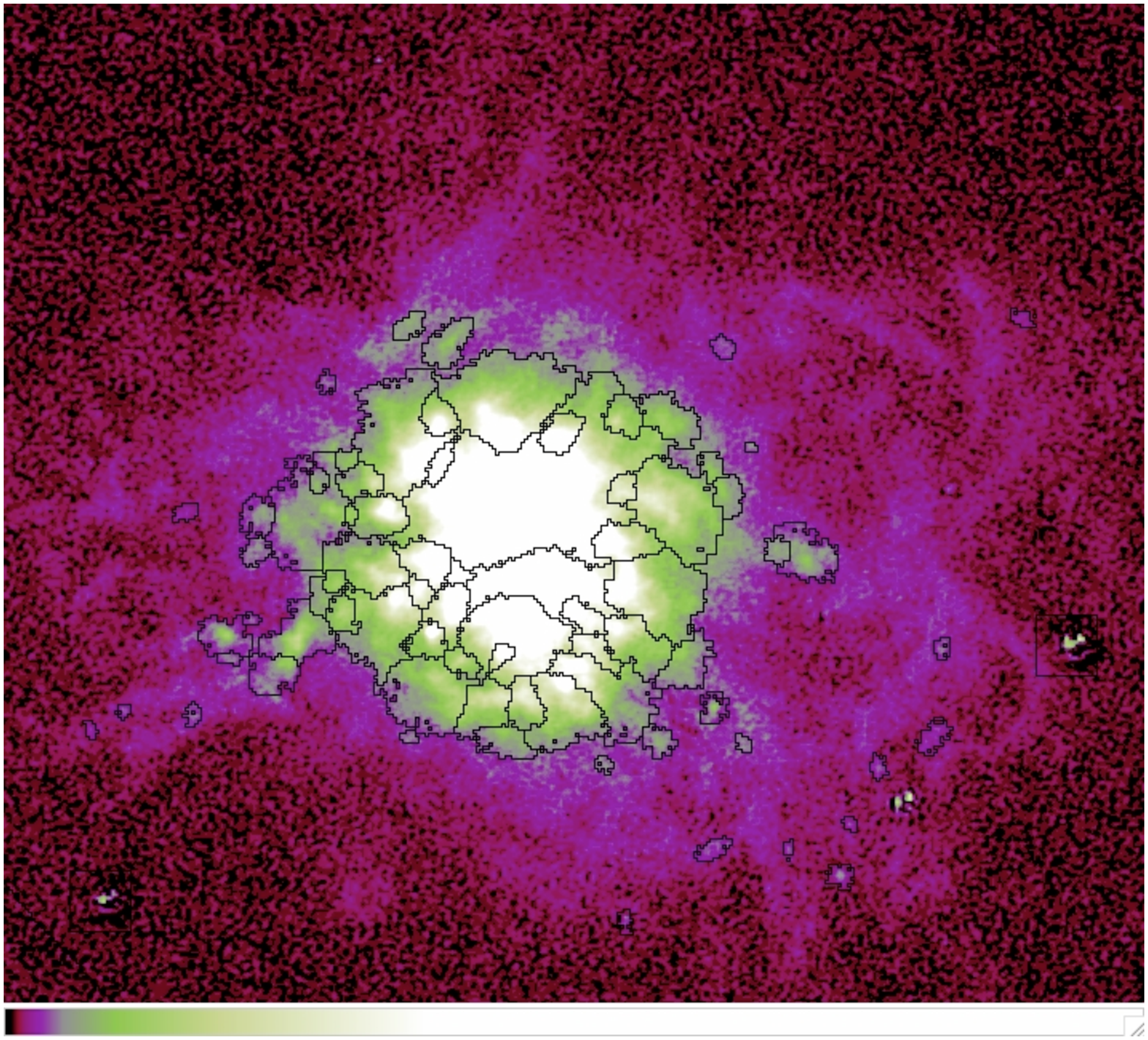}
\plotone{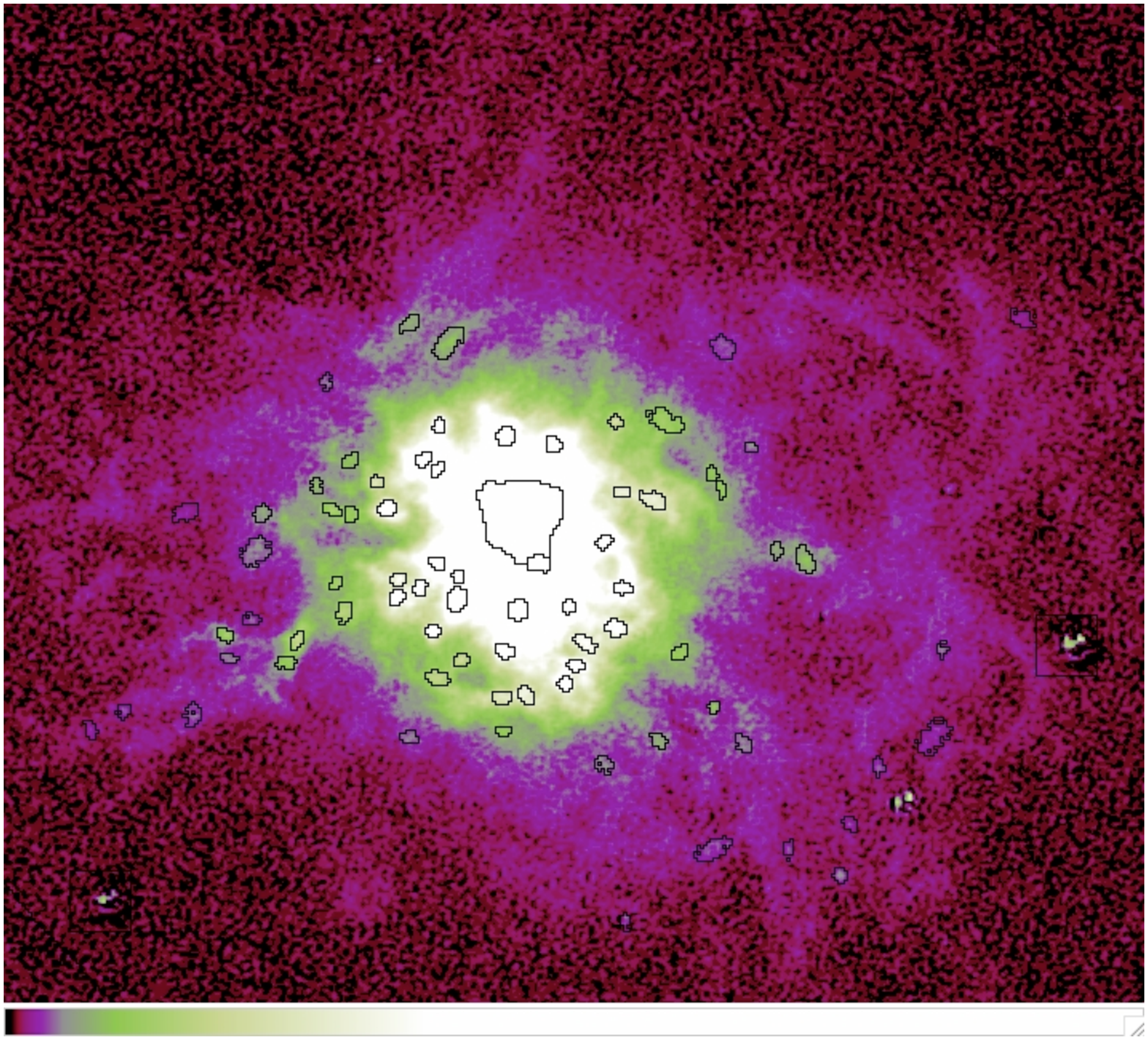}
\plotone{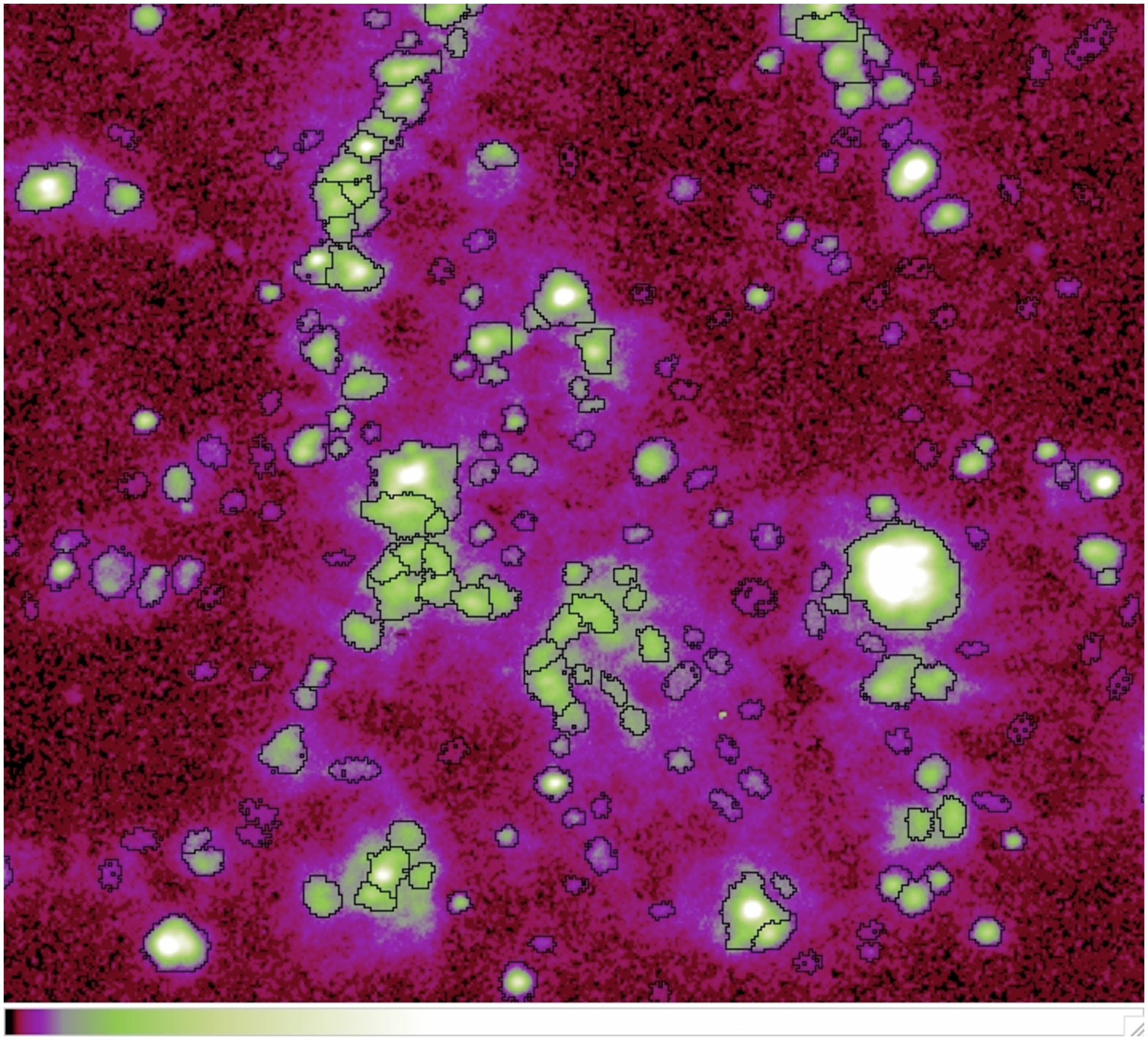}
% \plotone{J1339-31a.eps}
% \plotone{J1339-31e.eps}
% \plotone{J1318-21b.eps}
%\vspace*{-4.5in}
\caption{Figure~\ref{f_imgcompare}$a$ shows nebular boundaries for
   starburst galaxy J1339-31A for the standard run (termgrad = 7.5 $\rm
   pc\ cm^{-6}\ pc^{-1}$, \fwim\ = 0.11); Figure~\ref{f_imgcompare}$b$
   shows J1339-31A for a run that forces \fwim\ to a typical sample
   value (termgrad = 500 $\rm pc\ cm^{-6}\  pc^{-1}$, \fwim = 0.54);
   and Figure~\ref{f_imgcompare}$c$ shows the standard run for a
   normal galaxy having a similar distance, J1318-21 (termgrad = 5
   $\rm pc\ cm^{-6}\  pc^{-1}$, \fwim = 0.60).  All images are
   displayed with the same scales and color table.  [This figure is
   displayed in color in the on-line edition.]
\label{f_imgcompare} }
\end{figure*}

Since the crowding of
\hii\ regions at high inclinations mimics the crowding seen in
starburst galaxies, the normal \fwim\ values measured for the edge-on
galaxies, in contrast to the starbursts, is also strong evidence of a
real reduction in \fwim\ for the starbursts.  In addition, note that
the contaminating [\ion{N}{2}] emission is on the order of 3
times higher in the WIM than in the \hii\ regions (e.g., Hoopes \&
Walterbos 2003; Collins \& Rand 2001; Shopbell \& Bland-Hawthorn
1998).  Thus, the true ratio of 
\Ha\ emission $L_{\ha}({\rm WIM})/L_{\ha}(\rm HII)$ is less than the
apparent relative fluxes observed in the \Ha\ filters.  Furthermore,
Figures~\ref{f_imgcompare}$a$ and $b$ compare, respectively, the
\hiiphot\ boundaries for the same starburst galaxy for the standard
run, and a run for which the output has been forced to match a value
of \fwim\ typical of the entire sample.  For the latter, we obtain
\fwim$=0.54$ with an adopted terminal gradient that is over 50 times
its value for the standard run.
Figure~\ref{f_imgcompare}$c$ shows another standard run for a normal
galaxy having a similar distance.  It is apparent that a gross
difference in nebular boundary criteria is necessary to force the
starbursts to match the normal galaxies in \fwim.
Thus, our result of a lower fraction of WIM emission from starbursts
appears to be robust, although we do caution that
specific, quantitative measurements of \fwim\ are dependent on the
method used to define and distinguish diffuse \Ha\ emission from \hii\
regions; readers should bear this in mind when comparing results
between different studies.

\section{Relation of WIM to Galaxy Properties}

Table~1 lists the SINGG SR1 sample of galaxies.\symbolfootnote[2]{We
  omit the multiple source J0514--61 owing to processing problems, and
  so our sample has only 109 galaxies instead of the full
  SR1 sample.}  For convenience, we
repeat several fundamental parameters from Paper~I here.  As detailed in
Paper~I, \hi-related\ quantities were derived from \hipass\ data, in
particular, {\sc HiCAT}, the full \hipass\ catalog (Meyer, et al. 2004),
and the \hipass\ Bright Galaxy Catalog (Koribalski et al. 2004).
Distances were derived primarily from the \hi\ radial velocities using
the distance model of Mould et al. (2000), with optical distances to
the nearest sources taken from the Catalog of Neighboring Galaxies
(Karachentsev et al. 2004).  Readers are referred to Paper~I for
further details regarding SINGG catalogued quantities.  The columns in
Table~1 are as follows:  (1) {\sc HiPASS} designation, (2) optical 
identification, (3) Hubble type, (4) distance, (5) inclination angle
$i$, (6) \hiiphot\ terminal gradient, (7) \hi\ mass \mhi, 
%  (8) Galactic \Ha\ extinction $A_{\rm H\alpha,gal}$, (9) internal \Ha\ extinction
% $A_{\rm H\alpha,int}$, 
(8) $R$-band luminosity $L_R$, (9) \Ha\
luminosity $L_{\ha}$, (10) \Ha\ surface brightness $\Sigha$, (11) \Ha\
diffuse fraction \fwim, and (12) critical star-formation rate (see below).
% (13) slope $\alpha$ of the low-luminosity end of the $\Sigha$
% distribution, and (14) standard error on $\alpha$.  
All quantities are in the SINGG database and presented by Paper~I and Hanish
et al. (2006), except for columns 6, and 9 -- 14, which are derived in this
work.  Our \Ha\ luminosities are corrected for Galactic and internal
extinction using the same values as those in Paper~I.

\begin{figure*}
\epsscale{1.3}
%\vspace*{2.0in}
%\hspace*{-3.5in}
\plotone{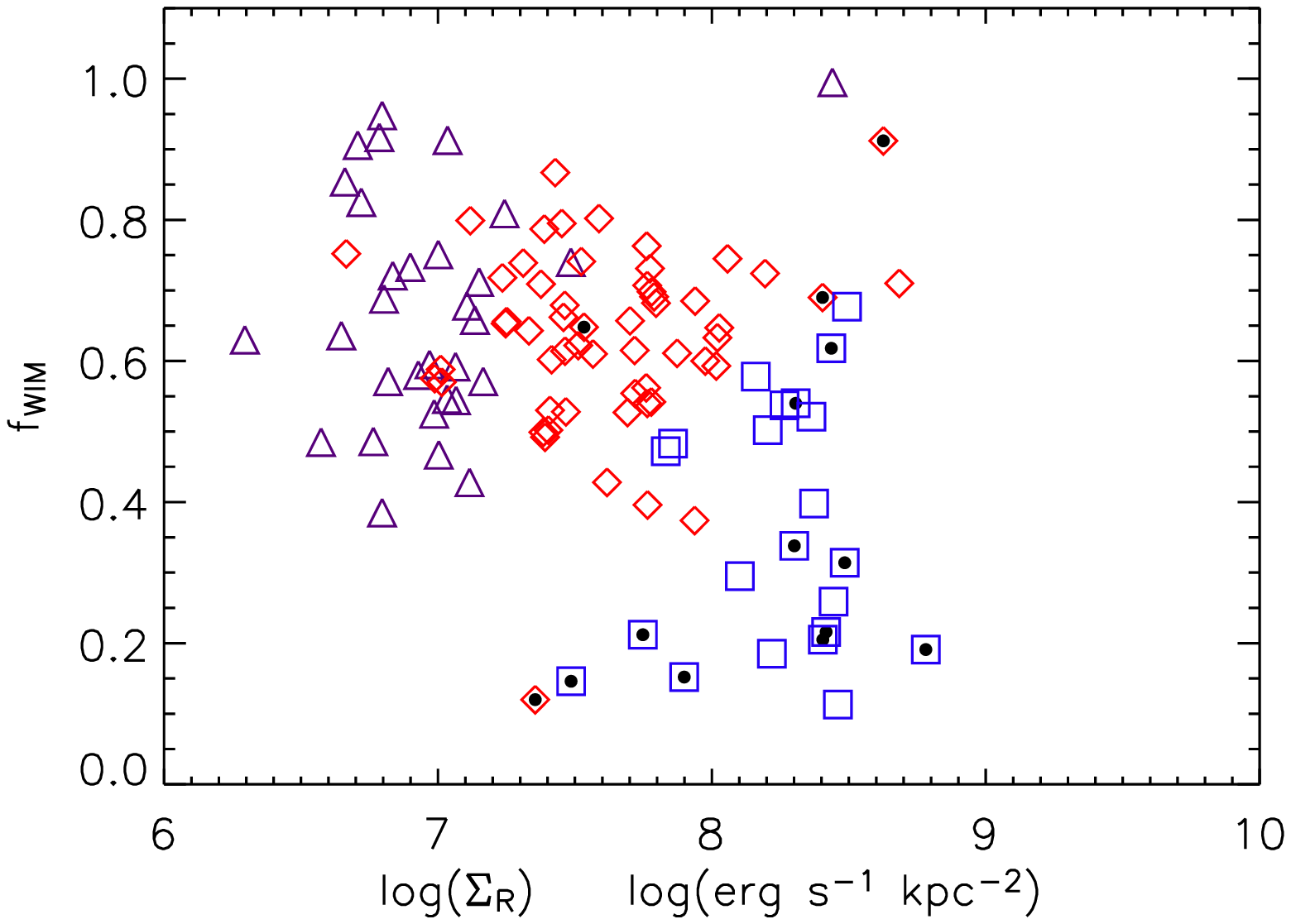}
% \plotone{diff_sb.ps}
%\vspace*{-4.5in}
\caption{\fwim\ as a function of $R$ surface brightness.  Symbols
  are as in Figure~\ref{f_termgrads}.
\label{f_fwimsb} }
\end{figure*}

We generally confirm earlier findings that the fraction of \Ha\
luminosity due to the WIM is largely 
independent of galaxy type, with the exception of the starburst
galaxies, as described above.  There is also a general trend of
decreasing \fwim\ with increasing $R$ surface brightness
(Figure~\ref{f_fwimsb}).  Helmboldt et al. (2005) and O'Neil
et al. (2006) also report that low surface-brightness galaxies tend
toward higher diffuse fractions.  Here, we see a much larger scatter
in \fwim\ for low surface-brightness galaxies.  This scatter 
may be due to uncertainties in continuum subtraction, which is more
important in these galaxies.  It is apparently not caused by
variations in the \hiiphot\ terminal gradient parameter, when
comparing outliers with typical galaxies.  For the 109
\hi-selected galaxies in SINGG SR1, the mean diffuse fraction is
0.59$\pm 0.19$.  This value is somewhat higher than the typical
$\sim$40\% value found for other nearby galaxies 
(e.g., Thilker et al. 2002; Zurita et al. 2000; Ferguson et al. 1996).
Since most of the previous studies favored nearby, high
surface-brightness, actively star-forming galaxies, 
we believe that some of this difference is due to this optical
selection effect.  Furthermore, most of those nearby galaxies have large
angular extents over several arcminutes, and the galaxy apertures used
for the total \Ha\ luminosity measurements were generally 
smaller in physical size than those in the SINGG survey (see Paper~I).  We
note that Helmboldt et al. (2005) similarly studied an \hi-selected
sample and found typical \fwim $\sim 0.45$; however, their \fwim\
were measured within the $R$-band half-light radii.  Furthermore, the
absolute measured value of \fwim\ also likely depends on the depth of
the \Ha\ observations, since fainter diffuse emission may be detected
with deeper imaging.  The SINGG survey typically has sensitivity to an
emission measure of 1.4 pc cm$^{-6}$, or $2.21\times 10^{-17}\ \rm erg\
s^{-1}\ cm^{-2}\ arcsec^{-2}$ (Paper~I), which is deeper than
many earlier studies, although stated depth limits depend on factors
like background subtraction for any such work.  All of the above effects
are likely contributors 
to our higher mean \fwim\ value for the SR1 sample, and further detailed
investigation is necessary to evaluate these effects.  There
may also be some systematic measurement bias due to our specific
definition and algorithm for identifying the WIM relative to other
studies, as discussed above.  However, we emphasize that this is the
largest uniform study of the WIM to date, and it also benefits from
the lack of any optical biases in sample selection.

\begin{figure*}
%\epsscale{1.0}
%\vspace*{2.0in}
%\hspace*{-3.5in}
\plotone{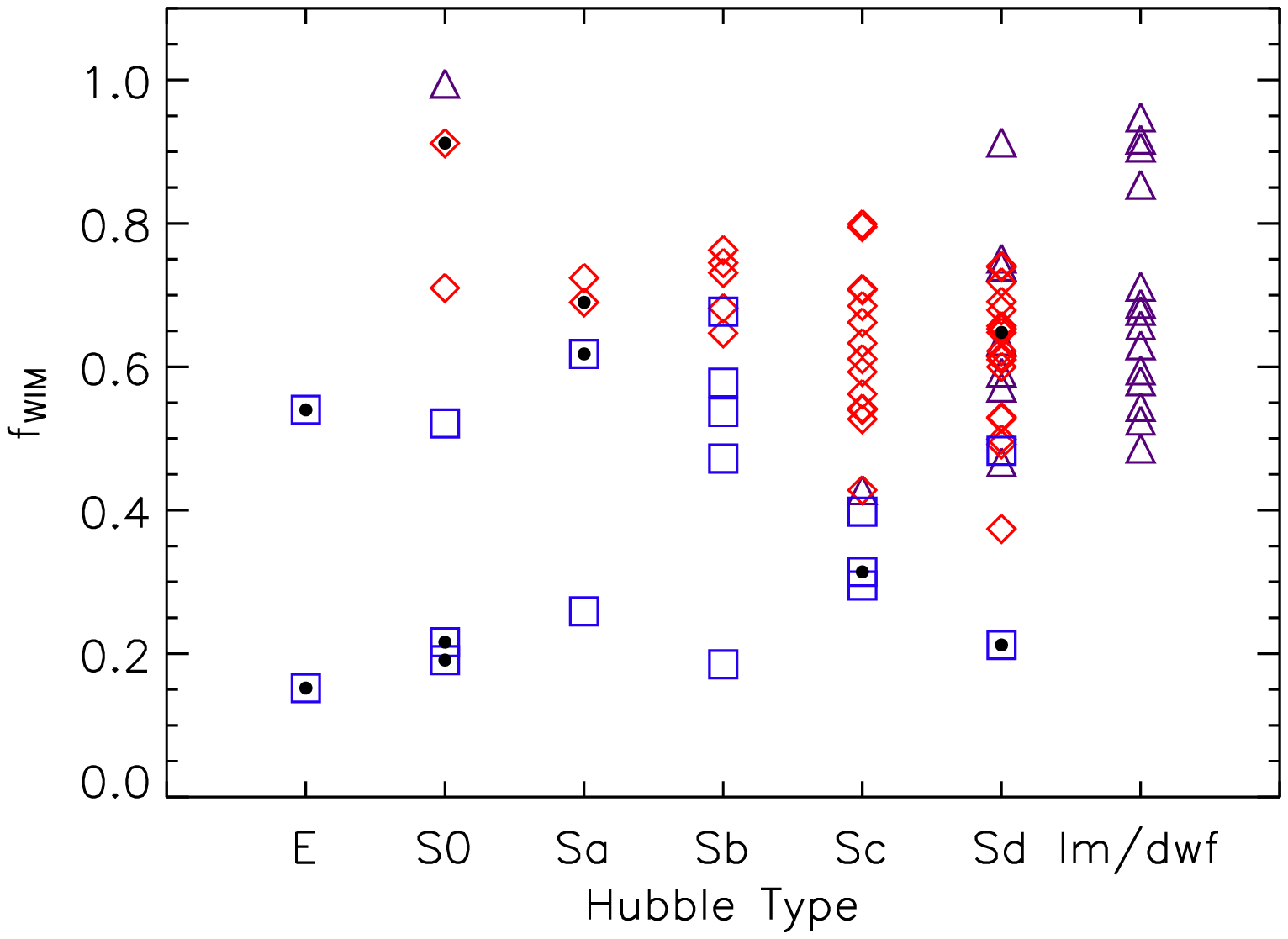}
% \plotone{diff_hub.ps}
%\vspace*{-4.5in}
\caption{WIM fraction for galaxy Hubble types.  Symbols for different
  star-formation categories are as in Figure~\ref{f_termgrads}.
\label{f_df_hubble} }
\end{figure*}

Figure~\ref{f_df_hubble} shows \fwim\ for the different
Hubble types.  Intermediate types are assigned to the
earlier type, for example, Sbc galaxies are binned with Sb galaxies.
Sm types are binned with Sd galaxies, and barred and non-barred types
are binned together.  The Hubble types are simply
those from the NED\symbolfootnote[3]{The NASA/IPAC Extragalactic Database
is operated by the Jet Propulsion Laboratory, California Institute of
Technology, under contract with NASA.} database.  In all of our 
analyses, including Figure~\ref{f_df_hubble}, we see that the nuclear 
starbursts tend to be less confined to the parameter space occupied by the
rest of the sample.  Apparently, nuclear starbursts take place in many
different galaxy conditions, and thus the star formation represented
by such events is not closely related to non-nuclear star-formation
properties, the latter of which are linked to the parent galaxies'
global parameters. 

\setcounter{table}{1}
Considering the remainder of the galaxies in Figure~\ref{f_df_hubble},
there appears to be no correlation in the mean \fwim\ with Hubble
type, although the scatter in \fwim\ tends to increase for late types.
Table~\ref{t_hubble} gives the means and standard deviations for
different galaxy classes, along with the number of galaxies $N$,
averaged in each class.  We again see that starbursts have
systematically lower diffuse fractions across all galaxy types.  We
discuss this further below.

\begin{deluxetable}{lccr}
\footnotesize
\tablewidth{0pt}
\tablecolumns{4} 
\tablecaption{Mean WIM fractions \label{t_hubble}} 
\tablehead{ \colhead{Category} & \colhead{$<f_{\rm WIM}>$} & \colhead{std
    dev } & $N$
 }
\startdata
All SR1 & 0.59 & 0.19 & 109 \\
E, S0, Sa\tablenotemark{a}& 0.54 & 0.29 & 12 \\
Sb & 0.60 & 0.17 & 10 \\
Sc & 0.57 & 0.15 & 18 \\
Sd/Sm & 0.61 & 0.14 & 26 \\
Im/dwf & 0.65 & 0.18 & 24 \\
$\log\Sigha \leq 38.4$\tablenotemark{b} & 0.67 & 0.16 & 31 \\
$38.4 < \log\Sigha \leq 39.4$\tablenotemark{b} & 0.63 & 0.13 & 56 \\
$39.4 < \log\Sigha$\tablenotemark{b} & 0.36 & 0.18 & 22 \\
Nuclear & 0.38 & 0.25 & 14 \\
\enddata
\tablenotetext{a}{Most of the E, S0, and Sa galaxies in this sample are
  dominated by nuclear star formation (see Figure~\ref{f_df_hubble}).}
\tablenotetext{b}{$\Sigha$ in units of $\rm erg\ s^{-1}\ kpc^{-2}$.}
\end{deluxetable}

\begin{figure*}[tpbh]
%\epsscale{1.0}
%\vspace*{2.0in}
%\hspace*{-3.5in}
\plotone{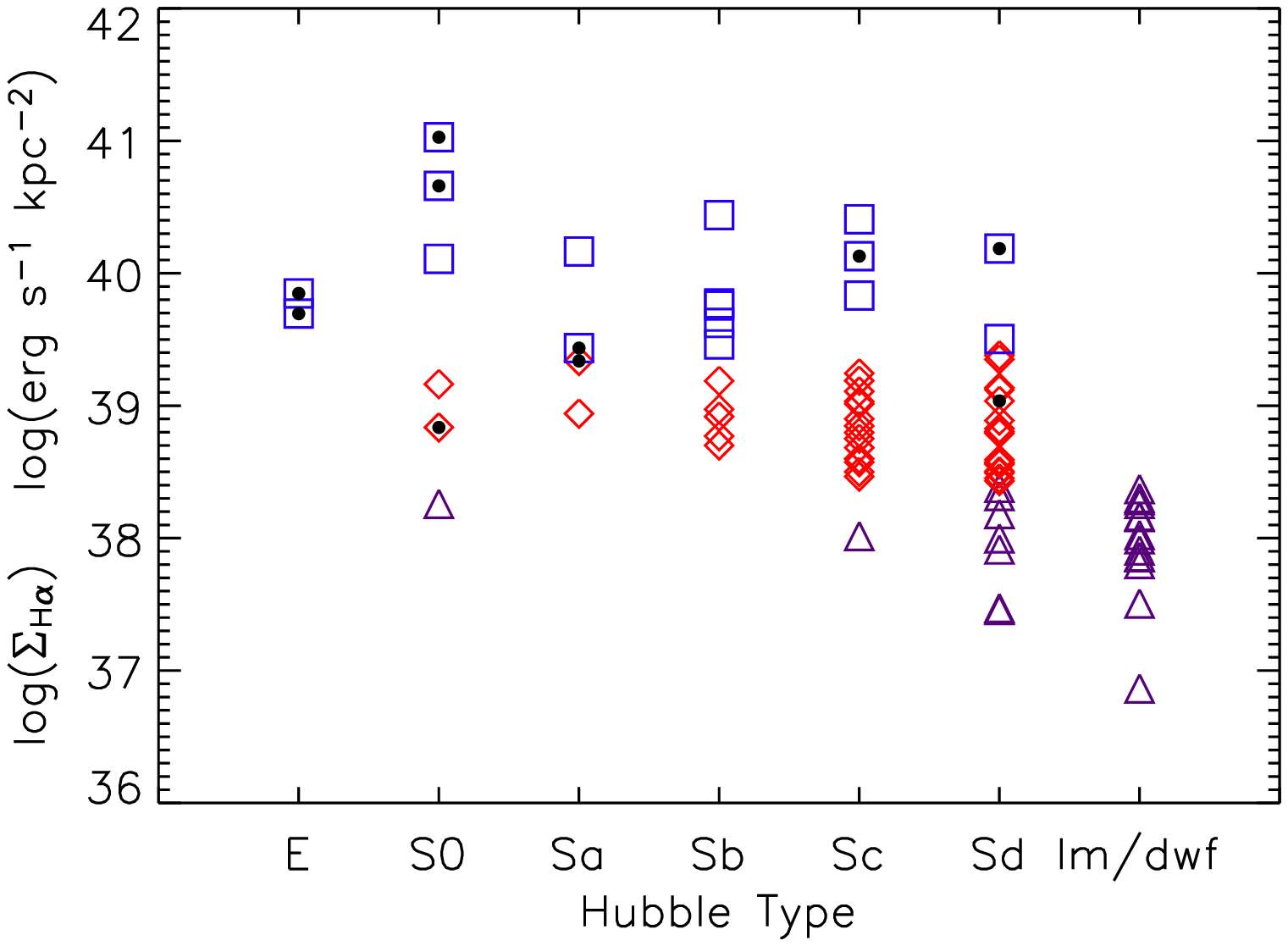}
% \plotone{SFIha_hub.ps}
%\vspace*{-4.5in}
\caption{\Ha\ surface brightness, a measure of star formation
  intensity, for different galaxy Hubble types.  Symbols are as in
  Figure~\ref{f_termgrads}. 
\label{f_sfi_hubble} }
\end{figure*}

% We might expect the \Ha\ diffuse fraction to be related to the star
% formation per unit area, or star formation intensity (equation~\ref{eq_sfi}).
Figure~\ref{f_sfi_hubble} shows the \Ha\ surface brightness vs Hubble
type for the sample.  
Although late-type galaxies are often said to have the highest SFR per
unit area, we see that the mean SFI is highest for Sb galaxies, again
with the scatter increasing with later Hubble type. 
Kennicutt (1998) notes that late-type galaxies tend to have more
extended star-forming disks, and that characterizations across the
Hubble sequence are dependent on how star formation is parameterized
relative to the parent galaxy properties.  He shows that it is the
global \Ha\ equivalent width that systematically increases toward
later galaxy types, which is also seen in our data.

\begin{figure*}
%\epsscale{1.0}
%\vspace*{2.0in}
%\hspace*{-3.5in}
\plotone{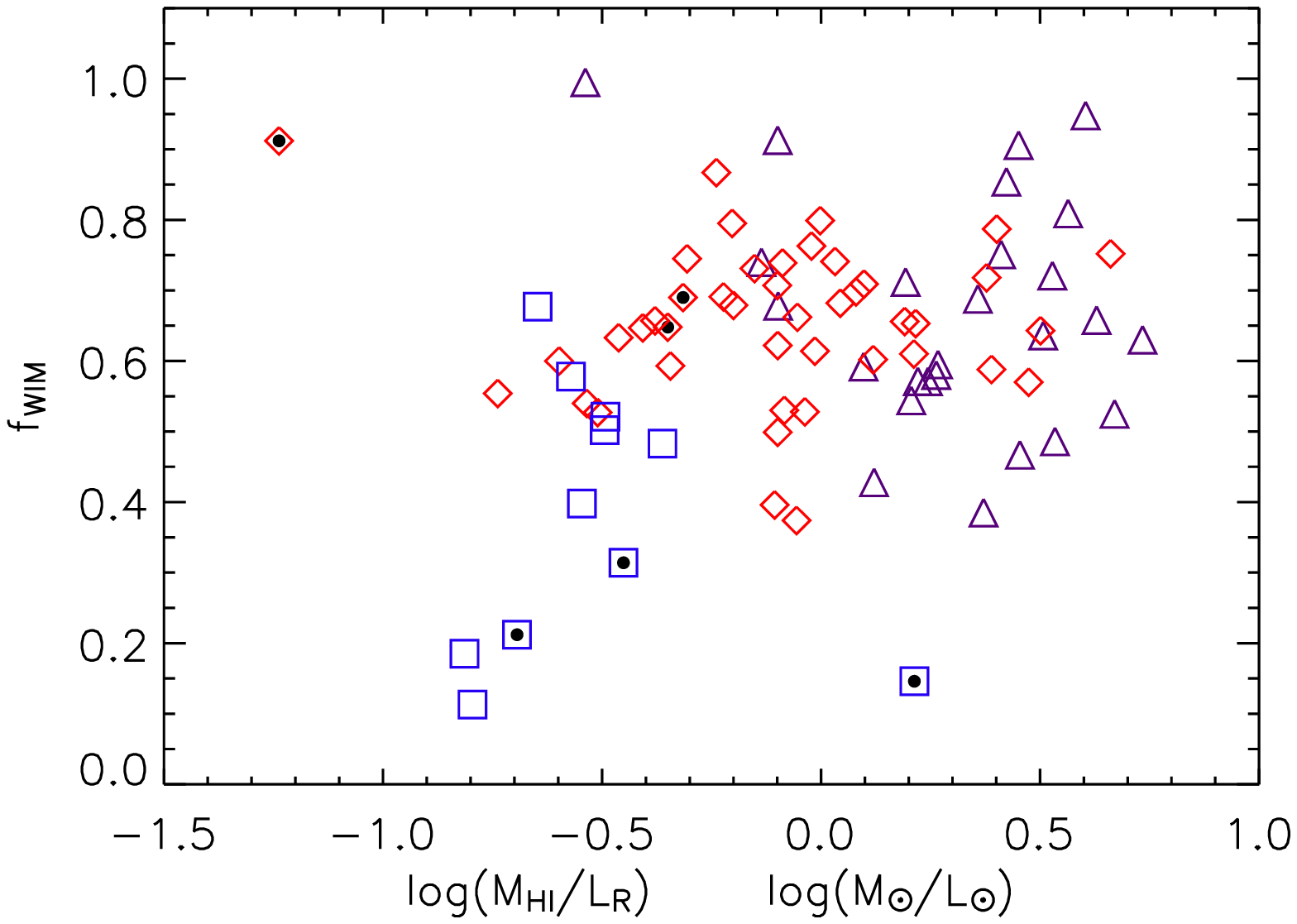}
% \plotone{diff_MHI.ps}
%\vspace*{-4.5in}
\caption{\Ha\ diffuse fraction vs galaxy {\sc Hi} gas fraction as
  measured by \mhi$/L_R$.  Symbols for different
  star-formation categories are as in Figure~\ref{f_termgrads}, and
  galaxies which have only upper limits on \mhi\ are omitted for clarity.
% arrows show upper limits on the gas fraction for multiple galaxies
% corresponding to a single {\sc Hi} detection.
\label{f_df_hifrac} }
\end{figure*}

Since the WIM represents ionization of the neutral ISM, we might
expect to find a relationship between the diffuse \Ha\ fraction and
galaxy \hi\ content.  Figure~\ref{f_df_hifrac} shows 
\fwim\ vs \mhi$/L_R$, the \hi\ mass normalized by the $R$
luminosity, which is a measure of galaxy \hi\ gas fraction.  
Note that for 28 galaxies, there are only upper limits on \mhi, since
these are multiple objects encompassed within the target {\sc HiPASS} beams.
A large scatter in \fwim\ is evident in Figure~\ref{f_df_hifrac}.
There is no apparent relation between \fwim\ and neutral gas fraction,
except for a few galaxies having the lowest values of \mhi$/L_R$,
which also show the lowest diffuse fractions.  Closer inspection shows
that these are starburst galaxies, and again,
that {\it ordinary starbursts as a group show both lower diffuse fractions
and lower \hi\ gas fractions.}  

\begin{figure*}
%\epsscale{1.5}
%\vspace*{2.0in}
%\hspace*{-3.5in}
% \plotone{diff_SFIha.ps}
\plotone{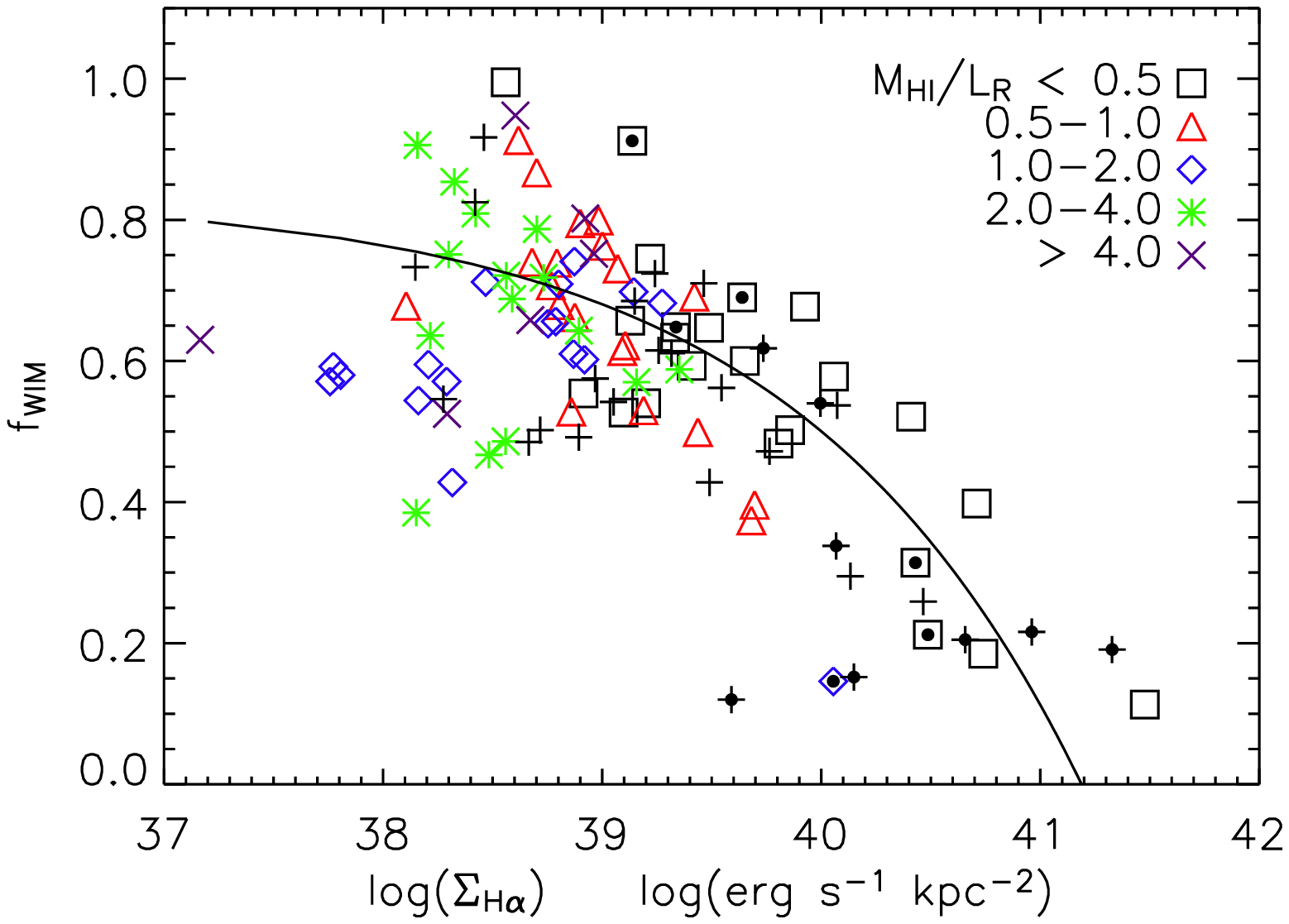}
% \plotone{diff_SFIha_MHI.ps}
%\vspace*{-4.5in}
\caption{\Ha\ diffuse fraction vs \Ha\ surface brightness, a
  measure of star formation intensity.  
% Figure~\ref{f_df_SFI}$a$ has symbols for different star-formation categories as in
%  Figure~\ref{f_termgrads}.  
  Symbols correspond to the {\sc Hi} gas fraction \mhi/$L_R$ as shown,
  in units of $\msol/L_\odot$, and plus symbols indicate galaxies
  whose {\sc Hi} gas fractions are only upper limits.  
  Note that starburst galaxies are defined as those having
  $\log\Sigha/(\rm erg\ s^{-1}\ kpc^{-2}) > 39.4$.  The curve has a
  form given by equation~\ref{eq_densbound}.
\label{f_df_SFI} }
\end{figure*}

Figure~\ref{f_df_SFI} shows the \Ha\ diffuse fraction vs star
formation intensity, with symbols now corresponding to \hi\ gas fraction, 
\mhi/$L_R$.  We see an anti-correlation that is largely defined by the
starbursts at $\log\Sigha> 39.4$; the non-starburst galaxies show
little, if any, trend in \fwim\ with $\Sigha$.
% This figure omits the nuclear starbursts, which, as we have seen, have star
% formation properties that are essentially independent of the the
% global galaxy properties that are related to the \Ha\ emission for the
% rest of the sample. 
% As before, some galaxies are missing because they do
% not have available \hi\ mass measurements.  
Interestingly, Figure~\ref{f_df_SFI} shows that the galaxies with the {\it lowest}
\hi\ gas fractions define an upper edge in the \fwim\ -- SFI space,
whereas galaxies with the largest \mhi/$L_R$ have low star formation
intensities and wide range of \fwim.  There is a continuous relation
between \hi\ gas fraction, star formation intensity, and \fwim\ for
the galaxies near this edge.

\section{Escape of Ionizing Radiation?}

Why do the ordinary starbursts show both lower \Ha\ diffuse fractions
{\it and} lower \hi\ gas fractions?  Given the higher ionizing luminosities
in starbursts, the lower \hi\ fraction might not be surprising.
However, we might then have expected a {\it higher} \fwim\ if the lower
\hi\ fraction were due to ionization by the starbursts.  Since this is
not observed, we must consider other possibilities:  (1)  An important
ionization source of the WIM has been reduced, for example, ($1a$)
less ionizing radiation escaping from strongly obscured starbursts; or
($1b$) less ionizing radiation available from a reduced population of
field OB stars.  Alternatively, (2)  
% Most of the ISM is occupied by the starburst, leaving only a small,
% {\it density-bounded} fraction of diffuse ionized ISM.  In the latter
% case, we expect ionizing radiation to escape from such galaxies.
some fraction of ionizing radiation escapes from these galaxies,
either ($2a$) under-ionizing the WIM, or ($2b$) fully ionizing it in a
density-bounded situation. 

While we have no data at present to evaluate possibility (1$a$) above,
we can examine possibility (1$b$) by assuming simple global
parameterizations for the behavior of massive stars in the field.
Recent work suggests that field O and B stars ionize about
half of the WIM in ordinary star-forming galaxies like M33 (Hoopes \&
Walterbos 2000) and the Small Magellanic Cloud (Oey et al. 2004).
Observations suggest that ionizing radiation escaping from ordinary
\hii\ regions can account for the other half of WIM ionization (Oey \&
Kennicutt 1997; Voges et al. 2005).
However, in galaxies with high absolute star-formation rates, the
fraction of field OB stars is expected to decrease as (Oey et al. 2004),
\begin{equation}
f_{\rm field}= \bigl(\ln N_{*,\rm up}+0.5772\bigr)^{-1} \quad , 
\end{equation}
where $N_{*,\rm up}$ is the number of OB stars in the richest cluster
of that galaxy.  This assumes a simple power-law
relation for the clustering law of the form, 
\begin{equation}
N(N_*)\ dN_* \propto N_*^{-2}\ dN_* \quad ,
\end{equation}
where $N_*$ is the number of OB stars per cluster.  This clustering law
appears to be universal (e.g., Elmegreen \& Efremov 1997) and
is supported by observations of super star clusters (Meurer et
al. 1995; Zhang et al. 2001) and ordinary \hii\ regions (Oey \&
Clarke 1998).  It also appears to extend down to the individual 
massive stars in the field (Oey et al. 2004).  Therefore, we do expect
the contribution of 
ionization from field stars to decrease from about one-quarter of the
total ionizing emission in ordinary star-forming galaxies to about
one-tenth in galaxies with the highest SFR.  Since our starburst
galaxies, defined as those with the highest SFI, may well have extreme
SFRs, a lower fraction of field massive stars is plausible.

\begin{figure*}
%\epsscale{1.0}
%\vspace*{2.0in}
%\hspace*{-3.5in}
\plotone{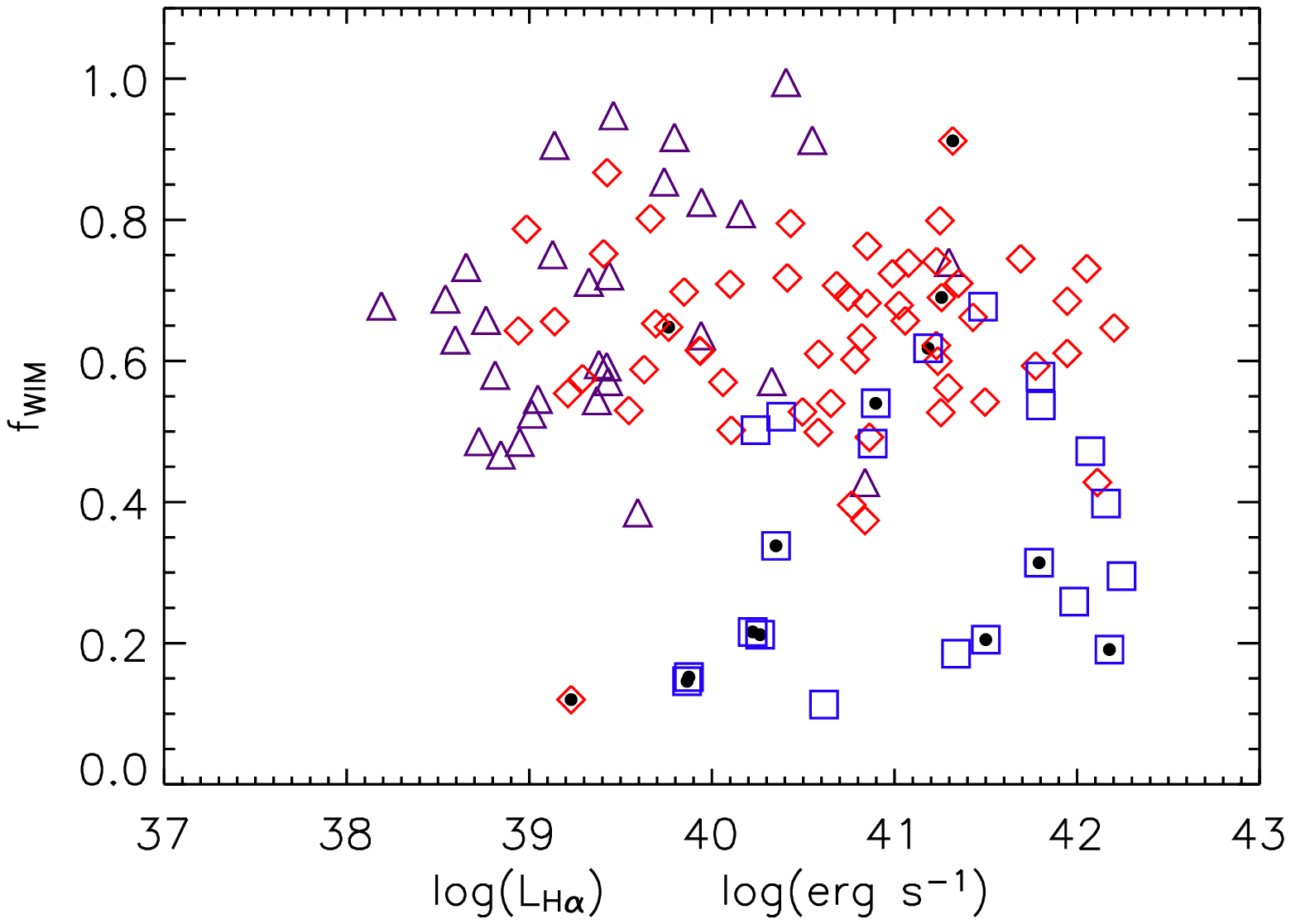}
% \plotone{diff_Lha.ps}
%\vspace*{-4.5in}
\caption{\fwim\ vs total star formation rate as measured by galaxy
  total \Ha\ luminosity.  Symbols for different
  star-formation categories are as in Figure~\ref{f_termgrads}.
\label{f_df_LHa} }
\end{figure*}

We test this possibility in Figure~\ref{f_df_LHa}, which shows the \Ha\
diffuse fraction vs total \Ha\ luminosity of the sample galaxies.  The
symbols correspond to the star formation categories as before.
Certainly there is no well-defined anti-correlation,
as would be expected if this scenario were the origin of that seen
in Figure~\ref{f_df_SFI}.  Thus, possibility ($1b$) above, that
starbursts have lower \fwim\ strictly because of a reduced population of
ionizing field OB stars, appears to be ruled out. 

We now consider possibility (2$a$) above, that ionizing radiation escapes
from the starbursts, under-ionizing the diffuse ISM.  Clarke \& Oey (2002)
calculated a crude parameterization for a threshold in star formation
rate above which galaxies are expected to release ionizing photons and
galactic superwinds:
\begin{equation}\label{eq_sfrcrit}
{\rm SFR_{crit}} = 0.15\Biggl(\frac{M_{\rm ISM,10}
  \tilde{v}_{10}^2}{f_d}\Biggr ) \ \ \rm M_\odot\ yr^{-1} \quad ,
\end{equation}
where $M_{\rm ISM,10}$ is the ISM mass in units of $10^{10}\ \rm
M_\odot$, $\tilde{v}_{10}$ is the thermal velocity of the ISM in units of
$10\ \kms$, and $f_d$ is a geometric correction factor for
disk galaxies.  This relation results from a simple criterion that
balances the supernova mechanical energy resulting from star formation
against the total ISM thermal energy.  If the former dominates, the
ISM is shredded, a galactic outflow is generated, and ionizing photons
escape.  For our SR1 sample, we adopt \mhi\ for the ISM mass.  The ISM
thermal velocity is roughly the sound speed, which we take to be 10
$\kms$ for the entire sample.  Similarly, we adopt $f_d = 0.1$ for the
entire sample, an approximate estimate for disk galaxies.  This value
for $f_d$ may be an underestimate for earlier types and irregular
galaxies, but given the uncertainty and crudeness of the relation in
equation~\ref{eq_sfrcrit}, we do not adjust $f_d$ for different galaxy
types.  Thus, we take SFR$_{\rm crit}$ to be simply proportional to
\mhi.

\begin{figure*}
%\epsscale{1.0}
%\vspace*{2.0in}
%\hspace*{-3.5in}
% \plotone{sfrcrita.ps}
\plotone{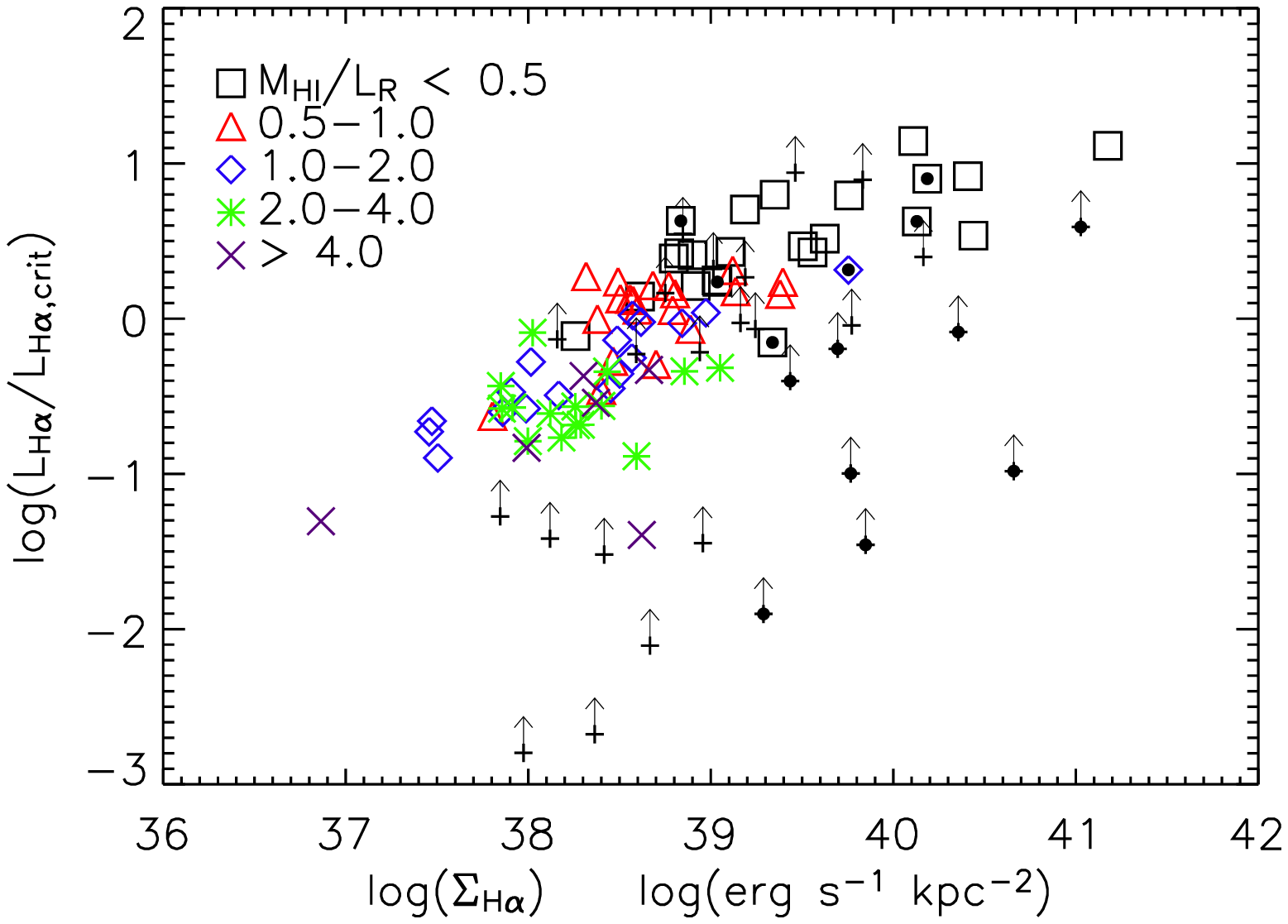}
% \plotone{sfrcritb.ps}
%\vspace*{-4.5in}
\caption{$\Lha/\Lhacrit$ vs \Ha\ surface brightness, a measure of star
  formation intensity.  Symbols show the {\sc Hi} gas fraction as
  in Figure~\ref{f_df_SFI}, and arrows indicate lower limits.  
\label{f_sfrcrit} }
\end{figure*}

\begin{figure*}[tpbh]
%\epsscale{1.0}
%\vspace*{2.0in}
%\hspace*{-3.5in}
\plotone{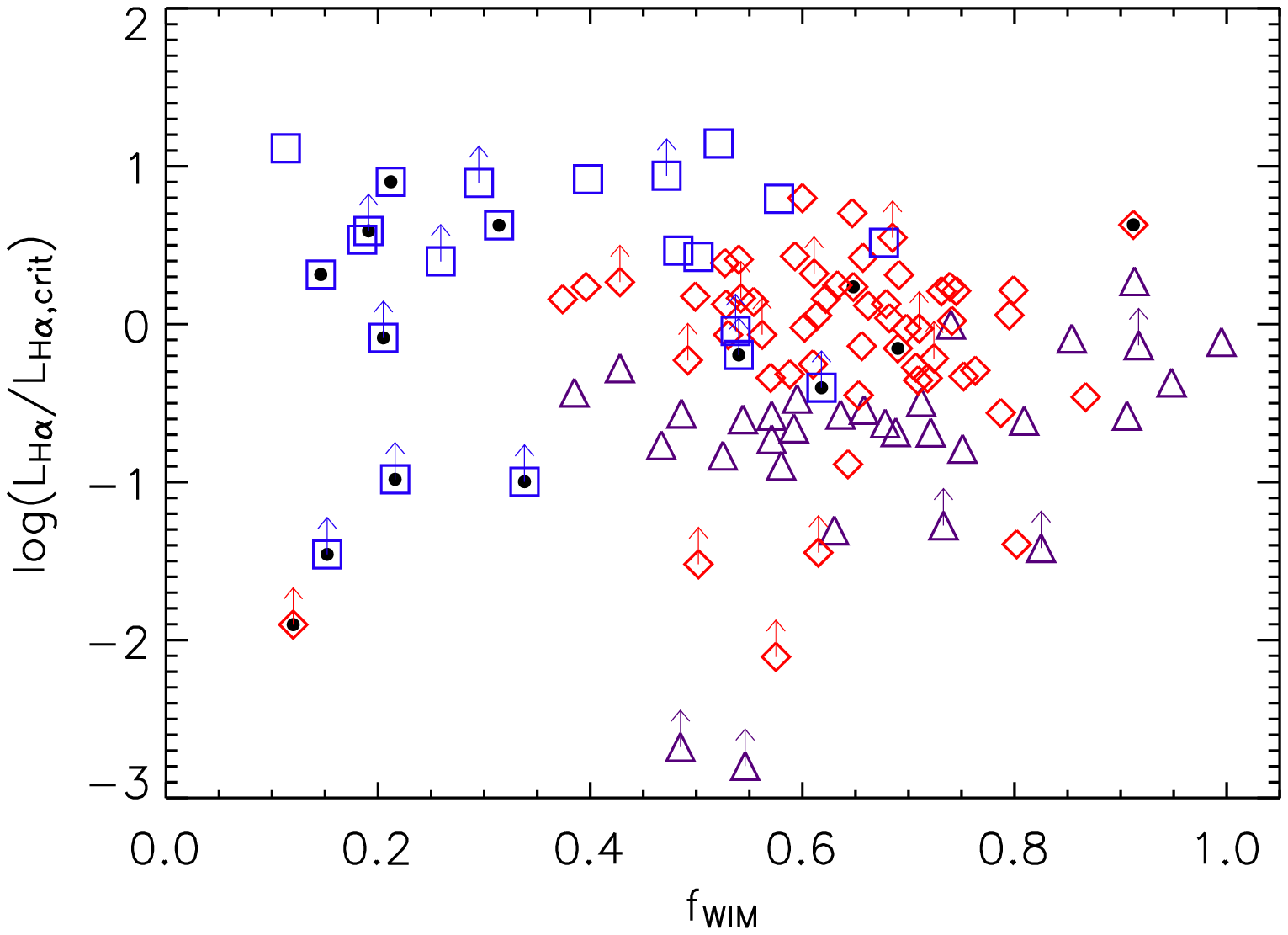}
% \plotone{sfrcrit_fwim.ps}
%\vspace*{-4.5in}
\caption{$\Lha/\Lhacrit$ vs \Ha\ diffuse fraction.
Symbols are as in Figure~\ref{f_termgrads}.
\label{f_sfrcrit_fwim} }
\end{figure*}

The \Ha\ luminosity is a measure of the total SFR, bearing in mind
that if ionizing photons are lost, then $\Lha$ is an underestimate of
the SFR.  We use the same relation as before (equation~\ref{eq_sfrha}) 
to estimate values for $\Lhacrit$ from the computed SFR$_{\rm crit}$.
In Figure~\ref{f_sfrcrit}, we show $\Lha/\Lhacrit$ vs SFI.
% As expected, there is a strong correlation between
% $\Lha/\Lhacrit$ and SFI.  
Roughly half the sample shows $\Lha/\Lhacrit
> 1$, suggesting that our crude relation estimating SFR$_{\rm crit}$
(equation~\ref{eq_sfrcrit}) may be subject to a scaling effect with a
factor of a few, but is not unreasonable.  Inclusion of molecular ISM,
for example, can be critical (e.g., Young \& Knezek 1989; Keres et
al. 2003) and would reduce the number of galaxies exceeding this
criterion.  The symbols in
Figure~\ref{f_sfrcrit} are plotted according to $M_{\rm HI}/L_R$, showing
that the galaxies with the lowest \hi\ gas fractions have the highest 
$\Lha/\Lhacrit$.  There is also a hint of flattening in the
correlation between $\Lha/\Lhacrit$ and SFI, which may simply be a
bias that underestimates the extinctions for starburst galaxies (see
Paper~I).  But if it is real, it may suggest that $\Lha$ underestimates
the star formation and ionizing radiation for the most extreme
objects.  Such a trend, along with the lower \hi\ gas fraction for the
starbursts is consistent with the escape of Lyman continuum radiation
from such galaxies.  However, regardless of any flattening,
Figure~\ref{f_sfrcrit} crudely confirms the plausibility that the
starburst galaxies have SFR $>$ SFR$_{\rm crit}$ for the escape of
ionizing radiation.

Figure~\ref{f_sfrcrit_fwim} shows $\Lha/\Lhacrit$
vs \fwim, plotted by SFI category.  The starburst galaxies, which
have low \fwim, have relatively higher $\Lha/\Lhacrit$ compared to
the remainder of the sample.  We note that although previous studies
reported that starburst galaxies are optically thick to ionizing
photons (e.g., Heckman et al. 2001; Leitherer et al. 1995), Shapley et
al. (2006) recently reported unambiguous detection of Lyman continuum
radiation from individual galaxies at $z\sim 3$, and
Bergvall et al. (2006) also detect Lyman continuum emission
from the blue compact dwarf Haro~11.  In addition, radiative feedback
from the Milky Way appears to be responsible for ionizing nearby
high velocity clouds (e.g., Bland-Hawthorn \& Maloney 1999; Putman et
al. 2003), which seems to be consistent with the prediction from
equation~\ref{eq_sfrcrit} (Clarke \& Oey 2002).  However, all of the
positive detections imply escape fractions of $\lesssim 0.1$.  

Figures~\ref{f_sfrcrit} and \ref{f_sfrcrit_fwim} show 
that the low fractions of diffuse \Ha\ emission in starburst galaxies are
qualitatively fully consistent with a predicted threshold SFR for
escaping ionizing radiation.  Indeed, these survey data suggest that
the crude prediction given by equation~\ref{eq_sfrcrit} is not only
meaningful, but also that we may be able to calibrate it with a
somewhat larger dataset, for example, the full SINGG sample.
Note that while equation~\ref{eq_sfrcrit} predicts the
loss of ionizing photons from galaxies, it is not based on an actual
density-bounding criterion caused directly by photoionization.
Thus it can straightforwardly account for the co-existence of some
neutral gas {\it and} substantial losses of ionizing radiation.  This
is fully consistent with the observations for our sample.  SFR$_{\rm crit}$
is instead based on an over-pressure criterion, in which star formation
drives the ISM pressure.  The higher temperatures determined for the
WIM in the starburst galaxy M82 by Shopbell \& Bland-Hawthorn (1998)
imply higher thermal pressures, supporting this scenario.  Our
over-pressure criterion is essentially
the same model considered by Wang, Heckman, \& Lehnert (1998) to
explain the anticorrelation of [\ion{S}{2}]/\Ha\ with SFI for actively
star-forming galaxies.  They separately considered ISM pressure
regulated by mechanical feedback and by simple hydrostatic
equilibrium.  The Clarke \& Oey (2002) relation
(equation~\ref{eq_sfrcrit}) considers both of these as components of 
a unified system, so our SINGG results also point to pressure regulation
of WIM and other feedback properties, like outflows and superwinds,
that are directly driven by star formation. 

Finally, we note possibility (2$b$) above, that the huge \hii\ regions
in starbursts have the appearance of almost fully occupying the ISM of
their host galaxies, suggestive that the diffuse WIM is the small,
remaining fraction in a density-bounded situation. 
For ionization-bounded conditions, a galaxy's
total Str\"omgren volume scales directly with the sum of all the
\hii\ region luminosities, assuming the escape fraction of photons
from \hii\ regions into the WIM remains constant.  Therefore, \fwim\
also remains constant, with the WIM again defined as the \Ha\ emission
exterior to the \hii\ regions.  However, for density-bounded
situations, a maximum possible Str\"omgren volume is reached, and thus
the WIM shrinks as the \hii\ regions grow.  If this is the cause of
our observed decrease in \fwim, then it should roughly follow a relation, 
\begin{equation}
f_{\rm WIM} \sim V_{\rm S,gal} - \sum{R_{{\sc Hii}}}^3 \quad ,
\end{equation}
where $V_{\rm S,gal}$ is the maximum possible Str\"omgren volume, and
$R_{{\sc Hii}}$ are the individual \hii\ region radii.  The
total volume in \hii\ regions scales directly with the SFR, while
$V_{\rm S,gal}$ is constant, so: 
\begin{equation}\label{eq_densbound}
f_{\rm WIM} \sim V_{\rm S,gal} - L_{\ha}^{1/3} \quad ,
\end{equation}
ignoring coefficients.  The curve overplotted in Figure~\ref{f_df_SFI}
shows this $-\Lha^{1/3}$ relation, scaled arbitrarily.  This ignores
the different galaxy sizes, which would introduce a dispersion in the
relation.  The curve shows a remarkable agreement with the data, and thus,
density-bounding in the 
inner volumes of galaxies is another possibility to explain the lower
\fwim\ in starbursts.  This would require the observed \hi\ to be in
the outer disks of such galaxies.  Putman et al. (2003) discuss these
geometric considerations concerning the escape of ionizing 
radiation from the Milky Way.

If the loss of ionizing radiation is indeed the origin of the low \Ha\
diffuse fraction in our starburst galaxies, then the implied fractions
of total Lyman continuum emission escaping into the IGM are
uncomfortably large, especially in view of the fact that we have no
truly extreme examples of starbursts in our sample.  If \fwim\ is
reduced by a factor of two or more, as seen in many objects, then the
implied escape fraction of ionizing radiation is on the order of
25\%.  This is much larger than fractions of $\lesssim 5$\% that have
been measured by direct, but very few, Lyman continuum observations.
It is therefore likely that more than one process is responsible for
the observed reduced \fwim\ in galaxies with the highest SFI.  For
example, possibility (1$a$) above remains:  that high dust content in 
starbursts is absorbing a disproportionate fraction of the Lyman
continuum emission.  It may also be, as suggested by Dopita et
al. (2006), that the WIM is composed largely of highly evolved,
filamentary \hii\ regions, and that these are fractionally
underrepresented in starbursts.  Further study is necessary to clarify
all of the possible models.  We note that if
ionizing radiation is indeed escaping from these galaxies, then models
suggesting that the bulk of \hi\ in galaxies is due to
photodissociation of large reservoirs of molecular gas (e.g., Allen
2001) are not supported by this result, since we see a decrease in
both \fwim\ and the \hi\ gas fraction in starbursts. 

\section{Conclusion}

We have used the SINGG SR1 dataset to study the properties of the
diffuse, warm ionized medium across the range of galaxy properties
represented in this \hi-selected sample.  The mean fraction of
diffuse \Ha\ emission from our galaxies is $0.59\pm 0.19$, somewhat higher
than found in previous studies.  We attribute much of this difference
to the lack of optical bias in our sample.  As with other studies, we
find no correlations in \fwim\ with Hubble type.  

However, starburst galaxies, defined here
as those having \Ha\ surface brightness $> 2.5\times 10^{39}\ \rm
erg\ s^{-1}\ kpc^{-2}$, show substantially lower \fwim\ compared to
other galaxies.  We caution that the magnitude of this result is
sensitive to the method of determining \fwim, but the \Ha\ surface
brightness distributions show it to be real.  We note that
[\ion{S}{2}]/\Ha\ and [\ion{N}{2}]/\Ha\ ratios, which are high in the
WIM, are similar between 
normal galaxies and starbursts (e.g., Lehnert \& Heckman 1994;
Kewley et al. 2001), consistent with \fwim\ in starbursts being
no larger than that in normal star-forming galaxies.  The effect of a
lower \fwim\ in starbursts does not appear to be dominated by a 
lower fraction of field OB stars.  However, it is broadly consistent
with the prediction that ionizing radiation is escaping from galaxies
having total star-formation rates above a critical threshold predicted
by Clarke \& Oey (2002).  This prediction derives from an ISM
over-pressure criterion, based on star formation driving the ISM
pressure and resulting WIM properties.  Wang et al. (1998) also
suggested that the WIM ionization state is determined by 
such a mechanism.  Supporting this
interpretation, we also find that galaxies with the highest
star-formation intensities tend to be those with the lowest \hi\ gas
fractions, suggesting that the gas has been consumed and/or ionized by
star-formation.  The data are also consistent with pure
density-bounding of the central regions in these galaxies.
If either of these models is correct, then it implies that ionizing
radiation is escaping from most starburst galaxies, with an inferred
escape fraction that may be as large as 25\% for this sample.
However, in view of the contradictory results in the literature that
suggest much lower escape fractions, it is likely that
other processes contribute to the reduction of the observed \fwim.

\acknowledgments
We thank Charles Hoopes and Karen O'Neil for useful discussions, and
Laura Woodney for assistance with IDL.  M.S.O., G.L.W., S.Y., and
S.M.C.-N. acknowledge support from the National Science Foundation,
grants AST-0204853 and AST-0448893.

%\vfill\eject
%|

\end{document}